\documentclass[aps,prb,twocolumn,letterpaper]{revtex4}
\usepackage{graphicx}
\usepackage{dcolumn}
\usepackage{amsmath}
\usepackage{amssymb}
\usepackage{verbatim}
\usepackage[caption=false]{subfig}
\usepackage{soul}
\usepackage[usenames,dvipsnames]{color}
\definecolor{orange}{rgb}{1,0.5,0}
\usepackage{float}
\usepackage{epstopdf}
\usepackage{ulem}
\def\kv{{\bf k}}

\def\Qv{{\bf Q}}

\def\beq{\begin{equation}}
\def\eeq{\end{equation}}

\def\beqa{\begin{eqnarray}}
\def\eeqa{\end{eqnarray}}

\begin{document}

\title{Entanglement Spectrum as a Probe for the Topology of a Spin-Orbit Coupled Superconductor}
\author{Jan Borchmann, Aaron Farrell, Shunji Matsuura and T. Pereg-Barnea}
\affiliation{Department of Physics and the Centre for Physics of Materials, McGill University, Montreal, Quebec,
Canada H3A 2T8}
\date{\today}
\begin{abstract}
The classification of electron systems according to their topology has been at the forefront of condensed matter research in recent years.  It has been found that systems of the same symmetry, previously thought of as equivalent, may in fact be distinguished by their topological properties.  Moreover, the non-trivial topology found in some insulators and superconductors has profound physical implications that can be observed experimentally and can potentially be used for applications.  However, characterizing a system's topology is not always a simple task, even for a theoretical model.  When translation and other symmetries are present in a quadratic model the topological invariants are readily defined and easily calculated in a variety of symmetry classes.  However, once interactions or disorder come into play the task becomes difficult, and in many cases prohibitively so.  The goal of this paper is to test whether the entanglement entropy and entanglement spectrum bare signatures of the system's topology. Using quadratic models of superconductors we demonstrate that these entanglement properties are sensitive to changes in topology.
\end{abstract}
\maketitle

\section{Introduction}
Over the past several years the study of topology in condensed matter systems has become a topic of great interest.  The topological properties of quantum Hall systems where studied since the 80's\cite{TKNN}, topological systems with time reversal symmetry were only predicted\cite{KaneMele2,KaneMele1} and realized\cite{Hsieh} recently.  The introduction of topology into the discussion of solid-state phenomena has revolutionized the classification of materials.  For instance, two insulating states in the same dimension and symmetry class, formerly thought of as being equivalent, could have a different topology and are not the same state of matter.  This classification is also supported by the direct physical implications of non-trivial topology, namely localized modes on system boundaries\cite{kitaev2,volovik,read}.  These modes are current carrying states on sample surfaces and Majorana fermions in vortex cores of topological superconductors.

In light of the above, it is desirable to assign a label which carries the information about the topology to any system of interest.  This is easy to do in a non-interacting system with translation invariance as it is described by a periodic, quadratic Hamiltonian.  This label is the `topological invariant', which is an integer number, related to Berry curvature in the Brillouin zone.  Loosely speaking, the invariant measures the phase winding of single particle states as the momentum is scanned in the Brillouin zone.  For example, in a two dimensional superconductor with broken time reversal symmetry, such as the model in the following discussion, the topological invariant is a Chern number, the integral of the Berry curvature over the Brillouin zone.  Calculating the Chern number requires knowing the single particle wave function at any point in the Brillouin zone and the presence of additional symmetries (like mirror or particle-hole) simplify the procedure greatly.

The ease with which one can evaluate the topological invariants in a non-interacting, clean system, unfortunately, does not carry over to dirty and/or interacting systems.  While breaking of translation invariance by disorder prevents the use of lattice momentum, interactions invalidate the notion of a single particle wave function altogether. K-theory classification\cite{schnyderPRB08,kitaev09,ryuNJP10} itself is valid in non-interacting dirty systems and there are formal ways of evaluating the topological invariants for interacting systems.  This could be based on flux insertion, similar to Laughlin's argument for quantum Hall systems\cite{KaneMele1} or using Green's functions\cite{volovik2, ghosh}.  However, these methods are not easy to implement, especially in situations where the ground state (only) is found numerically and given as a superposition of many configurations.
It is therefore desired to devise an alternative way of distinguishing a topological state from a trivial one in the presence of disorder and interactions and in a way that utilizes the ground state only, without requiring the full spectrum (or Green's function). For this reason we turn to study the entanglement entropy (EE) and entanglement  spectrum (ES).  We follow several authors who considered clean and non-interacting systems \cite{Oliveira, ding, bray-ali} and extend the study to other models of topological superconductors. Like previous authors we find that the topology is manifested in the entanglement properties in various ways.  We extend the previous studies and point to universal behavior that can be potentially used in more complicated cases\cite{Progress}.

The entanglement entropy and spectrum will be defined in the next section.  Before presenting the formal definitions, let us simply note that these include dividing the system to subsystems $A$ and $B$ and tracing out degrees of freedom associated with subsystem $B$.  Early applications of the entanglement entropy were concerned with how the entanglement entropy depends on the length of the boundary between the two subsystems. It was shown that, in two dimensions, if one considers a system with vanishing correlation length that the leading term in the entanglement entropy is linear\cite{kitaev,Levin2}, a property referred to as the {\it area law}.  Interestingly, in certain cases, the subleading length dependence of the entanglement entropy is directly related to topology.  This subleading term, $-\gamma$ is called the 'topological entanglement entropy'.  The cases for which $\gamma$ is non-zero are gapless topologically ordered states\cite{Levin1,Isakov} to be distinguished fromthe systems discussed in this work.

Although related to the topological ordering discussed above, topological insulators and topological superconductors are a different class.  A topological superconductor (or insulator) is a `symmetry protected topological state' (see for example Ref.~[\onlinecite{Turner}] for the distinction), meanwhile the topological order discussed above (where $\gamma$ is nonzero) is defined through long-range entanglement\cite{chen}. The bulk of a symmetry protected topological state is trivial from the point of view of topological order and therefore posses no topological entanglement entropy, i.e., $\gamma=0$. \footnote{The charged topological entanglement entropy \cite{ChargedSPT}, which is a universal sub-leading term of the charged entanglement entropy\cite{Belin:2013uta}, distinguishes symmetry protected topological phases from trivial phases.
On the other hand, as we show below, the leading term is sensitive to the topology, as might be expected, since the entanglement entropy is closely related to correlations.}

For the reason discussed above we cannot rely on $\gamma$ to distinguish between a topological and a trivial superconductors (or insulators). It is therefore natural to ask if the entanglement entropy contains any other signature that can be used to distinguish between a topological and a trivial phase in a symmetry protected topological state. Several proposals have been made over the past couple of years and we will outline the ones relevant to the current work. First, it has been shown that as one tunes model parameters across a topological phase transition a peak in the derivative of the entanglement entropy can be seen\cite{Oliveira}. This singularity occurs despite a lack of a rapidly changing $\gamma$, signalling the sensitivity of other terms in the entanglement entropy to the (symmetry protected) topology change. Second, one may also look for other subleading terms that contain potential information on the topology of the system. For example, there are logarithmic terms in the presence of corners and long range order\cite{metlitski}.

%edited below (Tami 22.10)
Signatures of the topology of symmetry protected states can also be found in the entanglement spectrum.  Again, the entanglement spectrum will be defined in the next section.  It is the spectrum of an auxiliary hamiltonian associated with the entanglement entropy.  The entanglement spectrum is sensitive to the type of partition applied.  For a partition whose boundary prevails throughout the whole sample, it has been shown that the low-lying entanglement spectrum mimics the excitation spectrum near a physical boundary, although the system may be fully periodic\cite{li, Qi}.  Moreover, Hsieh and Fu\cite{hsiehEE} showed that for an extensive partitioning the topological spectrum is related to bulk properties.  In this case, discussed below, a topological phase transition in the form of a gap closure can be seen in the entanglement spectrum by varying the partition only, even when the system parameters are unchanged.

%I commented out the next paragraph.  I think it belongs in the resubmission letter
%The motivation for this study is both practical and fundamental in origin. Fundamentally, the signatures of non-trivial topology in symmetry protected systems outlined above have only been applied to a handful of systems and in some cases ({\it }namely the partition induced gap closure) never to a superconductor. Thus by applying each of these methods to relevant superconducting models and comparing and contrasting with past results we feel that we are adding valuable additional information to the literature on this subject.  Practically, we are looking to develop a tool to apply to the a more difficult problem where the topology is not known {\it a priori}, for example to an interacting system. In this case it is crucial that we evaluate the potential use of each method in a more complicated scenario. To this end we need to explore non-topological, model dependent aspects of the analysis. For example, if a particular method is highly sensitive to things like finite size effects its potential use in the more complicated system will be very limited.  In addition to this, once we have the ability to apply any of these methods to an interacting system it will definitely be of interest to compare results to the non-interacting system.

% Deleted two red paragraphs here

%edited slightly below
The goal of this paper is to apply the tools outlined above to models of topological superconductors relevant for the search for Majorana fermions. In particular, the two models we consider describe spin-orbit coupled superconductors with various order parameters. We consider a superconductor with $d+id$-wave order parameter symmetry. This model is the mean field limit of an interaction driven superconductor with spin orbit coupling which has been studied previously by two of us\cite{Farrell1, Farrell2, Farrell3}. We also consider $s$-wave order parameter, motivated by recent proposals to realize topological superconductivity in heterostructures\cite{Sau, Alicea}.  In these proposals superconductivity is achieved by proximity to an $s$-wave superconductor.  It should be noted that in both the above models the pairing term is in the singlet s- or d-wave channel.  However, the presence of spin-orbit coupling forces the projection of this order parameter on to the spin-orbit coupled band.  In each band the projected order parameter acquires additional phase winding which alters the order parameter symmetry.  When there's only one relevant band the superconductor is topological with effective p- or f- symmetry.

%edited slightly below
The main findings of our study are as follows. (i) When varying model parameters such that the system changes its topology, the derivatives of the entanglement entropy with respect to model parameters are sharply peaked at the transition. This extends a previous observation by Oliveira\cite{Oliveira} to an additional system. This result holds even for very small subsystem sizes and could therefore find potential use in a more complicated system. (ii) Any effort to find subleading terms to the area law in our models using a 'finite' partition to $A$ and $B$ subsystems were overwhelmed by finite size effects. This is in contrast to previous studies\cite{ding, Oliveira} where the no such problems were reported.  Owing to this, studies based on subleading terms of more complicated systems may be of limited scope. However, by adopting a corner-less partition we were able to simulate large systems.  We establish that any possible subleading terms in the EE have to be due to corners\cite{papanikolaou}, as all sub-leading contributions are negligible in the corner-less partition.
Moreover, this shows that the slope in the area law is sensitive to the topology of the system and discontinuous at the phase boundary. Plotting this coefficient in parameter space is then a useful method for searching for a topological phase boundary. (iii) The entanglement spectrum of the corner-less partition provides a nice illustration of the connection between the low energy states in the entanglement spectrum of a partition with a prevailing edge and the low energy states of a physical system\cite{li, Qi}  with an edge for this model. (iv) In our model of a topological superconductor, a topological phase transition can be seen in the entanglement spectrum by varying the partition.  This result extends the work of Hsieh and Fu\cite{hsiehEE} on topological insulators to the case of topological superconductors and supports their general arguments.

%deleted a red paragraph and  commented paragraphs%

The rest of this Paper is organized as follows: In the next section we introduce the entanglement spectrum, entanglement entropy, and our model framework. In Section \ref{sec:PBEE} we present and discuss our results of the entanglement entropy in parameter space while in Section \ref{sec:area} we study the entanglement entropy as a function of system size.  Section \ref{sec:partition} contains our partition tuning study and concluding remarks are presented in Section \ref{sec:conclusion}.

\section{Model and Methods}

\subsection{The Reduced Density Matrix, Entanglement Spectrum and Entanglement Entropy}
We start by defining the reduced density matrix, the entanglement spectrum, and the entanglement entropy. We also discuss how they are obtained relatively simply in a non-interacting system. Starting from a ground state $|\psi\rangle$ one defines the reduced density matrix by dividing the system into two parts, A and B. The reduced density matrix\cite{li} of subsystem A is given by
\beq
\rho_A=\text{Tr}_B\left( |\psi\rangle \langle \psi| \right) \equiv \frac{e^{-H_A}}{ Z_A},
\eeq
where the trace is over all configurations of subsystem $B$ and the above equation serves as the definition of $H_{A}$, the entanglement Hamiltonian.  The entanglement spectrum is defined as the set of eigenvalues $\{E_i\}$ of the entanglement Hamiltonian, $H_A$.  $Z_A = \text{Tr}_A(e^{-H_A})$ is the partition function.
The entanglement entropy (EE) we choose to work with is the von-Neuman entropy, defined by:
\beq\label{eq:Sa}
S_A = \text{Tr}\left(\rho_A \log \rho_A\right)
\eeq
%
%As will be discussed below, a non-interacting fermionic system like ours leads to a quadratic $H_{A}$.  This allows for a significantly simplified procedure for calculating the reduced density matrix, $\rho_A$.  Once $\rho_A$ is obtained one can proceed to diagonalize it in order to find its eigenvalues and hence the eigenvalues $\{ E_{i}\}$ of the reduced Hamiltonian $H_A$.
%
We now specialize our discussion to the system at hand: a quadratic system with superconductivity. In order to calculate the ES we appeal to the fact that the entanglement spectrum of a quadratic system is completely determined by its correlations. To show this we generalize a method proposed in Refs.~[\onlinecite{peschel1, peschel2}]. We briefly review the main steps of the method here, adjusted to the case of a superconductor.
Consider a state $|\psi\rangle$ which is the ground state of some quadratic Hamiltonian.
%When calculating higher order expectation values of high order terms like $\langle c_{i}^\dagger c_{j} c_{\ell}^\dagger c_{n}\rangle$ Wick's theorem is obeyed and can be utilized.  In fact, Wick's theorem holds if and only if the Hamiltonian is quadratic\cite{PeskinSchroder}.\textcolor{red}{Is this true?}  Note that if we
%
$|\psi\rangle$ is a Slater determinant of single particle states and therefore obeys Wick's theorem.  Now let us
consider averages $C_{i,j} = \langle c_{i}^\dagger c_{j} \rangle$ where $i,j$ are both in subsystem $A$. This average must be completely determined by the reduced density matrix $\rho_A$. Moreover, since $|\psi\rangle$ is a determinant all averages must obey Wick's theorem.  Therefore for any local operator $\mathcal{O}_A$ in subsystem $A$, $\langle \mathcal{O}_A \rangle = \text{Tr}\left(\rho_A \mathcal{O}_A\right)$ and the trace must obey Wick's theorem. It follows that $\rho_A$ is an exponent of a quadratic entanglement Hamiltonian. Further, if $|\psi\rangle$ is a ground state with some pairing ({\it i.e.} a BCS like wave function) then the anomalous averages $\langle c_{i}^\dagger c^\dagger_{j} \rangle$ must be non-zero. From this it follows that $H_A$ must also contain pairing.

The considerations above lead us to write a general form for $H_A$ as follows
\beq
H_A = \sum_{i,j\in A}\left( c_i^\dagger h_{i,j} c_{j} +\frac{1}{2} \left(c_{i}^\dagger \Delta_{i,j} c_{j}^\dagger + \text{h.c.}\right)\right)
\eeq
where $i,j$ label both site and spin in subsystem $A$. The above Hamiltonian can be written as $H_A =   \psi^\dagger \mathcal{H} \psi$ where $\psi = (c_{1} ... c_{N} , c_{1}^\dagger..c_{N}^\dagger)^T$. The matrix $\mathcal{H}$ obeys particle-hole symmetry and thus it can be diagonalized as $\mathcal{H} = W D W^\dagger $ where $D= \text{diag}(E, -E)$ where $E = \text{diag}(E_1....E_N)$ with $E_i>0 \ \forall i$ and
\beq
W =  \left(  \begin{matrix}
      u & v^* \\
      v & u^* \\
   \end{matrix}\right).
\eeq
where $u$ and $v$ are matrices in position and spin space. If we now define the correlation matrix
\beq
G = \left(  \begin{matrix}
      \langle c_{i} c^\dagger_{j} \rangle &\langle c_{i}c_{j} \rangle  \\
      \langle c_{i}^\dagger c^\dagger_{j} \rangle & \langle c_{i}^\dagger c_{j} \rangle \\
   \end{matrix}\right)
\eeq
and calculate the averages in terms of traces over $\rho_A$, one can show that $G$ can be represented as $G = W \tilde{G} W^\dagger$ where $\tilde{G} = \text{diag}(I-f, f)$ with $f= \text{diag}(n_f(E_1)....n_f(E_N))$ with $n_f(x) = 1/(1+e^{x})$. We now make the observation that $G$ and $\mathcal{H}$ are diagonalized by the same transformation. Therefore if we define the first $N$ eigenvalues of $G$ as $\zeta_i = 1-f(E_i)$ then the entanglement spectrum is given by $E_i = \ln\left(\frac{\zeta_i}{1-\zeta_i}\right)$. Thus the entanglement spectrum is obtained via the following program. Using a ground state $|\psi\rangle$ we calculate $G_{i,j}$ for $i,j$ in subsystem A, diagonalize the matrix $G$ and then use its eigenvalues to obtain the entanglement spectrum.

Using the relation between the entanglement entropy and the entanglement Hamiltonian in Eq.~(\ref{eq:Sa}) and $\zeta_i = 1-f(E_i)$, we find
\begin{align}
\begin{split}
S_{A} = - \sum_{i}(\zeta_{i}\ln{\zeta_{i}} + (1-\zeta_{i})\ln{(1-\zeta_{i})}),
\end{split}
\end{align}
which is just the entropy of a free fermionic gas with energies $E_{i}$. For a vanishing correlation length, as expected for an insulator, the entropy has the form
\begin{align}
\begin{split}
S_{A} = \alpha L - \gamma + \mathcal{O}(1/L),
\end{split}
\end{align}
where $L$ is the length of the partition between the two sub-systems.  The first term, proportional to $L$ is referred to as the area law and the sub-leading term $\gamma$ is called the `topological entanglement entropy'\cite{kitaev,Levin2}.  This term only depends on the topology of the ground state and is thus universal. Since the entanglement Hamiltonian of a 2+1d topological system is related to the Hamiltonian of 1+1d conformal field theory\cite{kitaev}, one could obtain the above expression by taking the large $L$ limit of the CFT partition function.

For our bulk model we expect $\gamma$ to be zero\cite{ding,Oliveira} since our topological state is a symmetry protected one. The assumption of a vanishing correlation length $\xi$ is justified, as long as the characteristic length of each subsystem is large compared to $\xi$. Thus, for a general partition, this limit is inappropriate due to the presence of corners and one gets,
\begin{align}
\begin{split}
\alpha \rightarrow \alpha(\xi), \quad -\gamma \rightarrow - \gamma(L,\xi)
%\beta(L,\xi) - \gamma.
\end{split}
\end{align}
This can then lead to sub-leading terms in the entanglement entropy. In the following sections we show that in our systems any such $\gamma$ is associated with partition corners.

\subsection{Quadratic Hamiltonian with Pairing}

In our model we look at quadratic states with p-wave or f-wave pairing.  These pairing states are the result adding momentum-spin locking (via spin-orbit coupling) to systems which otherwise tend to pair in the singlet s- or d-wave channel\cite{Farrell1,Farrell2,Farrell3}.
These systems have translational invariance and can thus be diagonalized in momentum space and therefore their Chern number  (the relevant topological invariant) can be calculated exactly. This means, conveniently, that the topological phase diagram is known. We can therefore use this to analyze the results given by the entanglement spectrum and entanglement entropy.

The model we consider is as follows
\begin{equation}\label{MF}
H=T+H_{SO}+H_{SC},
\end{equation}
where,
\begin{equation}
T = -\sum_{\langle i,j\rangle,\sigma} c_{i,\sigma}^\dagger t_{i,j} c_{i,\sigma}
\end{equation}
is the tight binding kinetic energy where $t_{i,j}$ are the hopping amplitudes. Here we take $t_{i,j} = t_{i-j}$ and define its Fourier transform as $\epsilon_{\kv}$. For nearest neighbor hopping $\epsilon_{\kv}=-2t(\cos{k_x}+\cos{k_y})$. Next,
\beq
H_{SC} = \sum_{\kv}\left( c_{\kv,\uparrow} \Delta_{\kv} c_{-\kv,\downarrow} +\text{h.c.}\right),
\eeq
where $\Delta_{\kv}$ is the superconducting order parameter. In what follows when we refer to the $s$-wave model we mean an order parameter of the form $\Delta_\kv = \Delta_0$ while $d+id$-wave symmetry means we have used $\Delta_{\kv} = \Delta_1 (\cos(k_x)-\cos(k_y)) +i\Delta_2 \sin(k_x)\sin(k_y)$.  These electron pairing functions in the singlet channel transform into $p$- or $f$-wave functions when written in the spin-orbit coupled band basis.  This alone does not guarantee topological superconductivity as there is usually two bands with opposite chirality.  Therefore the condition for topological superconductivity is only one relevant band participating in the pairing
\cite{Alicea}. The spin-orbit coupling term takes the form
\begin{equation}
H_{SO} = \sum_{\kv} \Psi_{\kv}^\dagger \mathcal{H}_\kv \Psi_{\kv},
\end{equation}
where  $\Psi_{\kv}=(c_{\kv,\uparrow}, c_{\kv,\downarrow})^T$, $\mathcal{H}_\kv = {\bf d}_\kv \cdot \vec{ \sigma}$ (with $\vec\sigma$ a vector of Pauli matrices acting on the spin).  ${\bf d}_\kv$ could in principle take any form which is convenient to describe spin-orbit coupling. Here we choose ${\bf d}_\kv=(A\sin{k_x},A\sin{k_y},2B(\cos{k_x}+\cos{k_y}-2)+M)$ ($A,B$ and $M$ are material parameters which describe the various spin-orbit coupling and Zeeman strengths).  This choice resembles the spin-orbit coupling term used by Bernevig, Hughes and Zhang\cite{BHZ} in the description of 2d topological insulators.

The hamiltonian (\ref{MF}) satisfies
\beq
U_{C}H^{*}(-k)U^{-1}_{C}=-H(k),
\eeq
%\textcolor{red}{Added -1 in the superscript above}\\
where $U_{C}$ is a unitary operator $\sigma_y\otimes \mathbb{I}_2$ in the basis of $(\psi_{\kv},\psi^{\dagger}_{-\kv})$. Since $U^{*}_C U_C=-\mathbb{I}_{4}$, this topological superconductor belongs to Class C\cite{schnyderPRB08,kitaev09,ryuNJP10}.

One can block diagonalize this hamiltonian by a unitary transformation and the topological number is given by a doubled Chern number.
%The Chern number of this model can be calculated exactly in closed form.
Defining $\xi_{\kv} = \epsilon_\kv-\mu$ the Chern number is given by\cite{ghosh, Farrell1, Farrell2}
\beq
C_1 =  \frac{1}{i\pi} \log\left[\frac{Q(0,0) Q(\pi,\pi)}{Q(\pi,0)Q(0,\pi)}\right],
\eeq
where $Q(\kv) = \text{sgn} ( |\Delta_{\kv}|^2+\xi_\kv^2-{\bf d}_\kv^2)$.  For our particular model we have $Q(0,\pi)=Q(\pi,0)$, regardless of parameters. We are therefore left with
\beq\label{chern1}
C_1 =  \frac{1}{i\pi} \log\left(\text{sgn}\left[( |\Delta_{0}|^2+\xi_0^2-{\bf d}_0^2)( |\Delta_{\Qv}|^2+\xi_\Qv^2-{\bf d}_\Qv^2)\right]\right),
\eeq
where $\Qv=(\pi,\pi)$. Using the above formulation we can map the topological phase diagram of the superconductor described by the Hamiltonian $H$.

\section{Topological Phase Boundary and the Entanglement Entropy}\label{sec:PBEE}

\begin{figure}[t]
  \setlength{\unitlength}{1mm}

   \includegraphics[scale=.5]{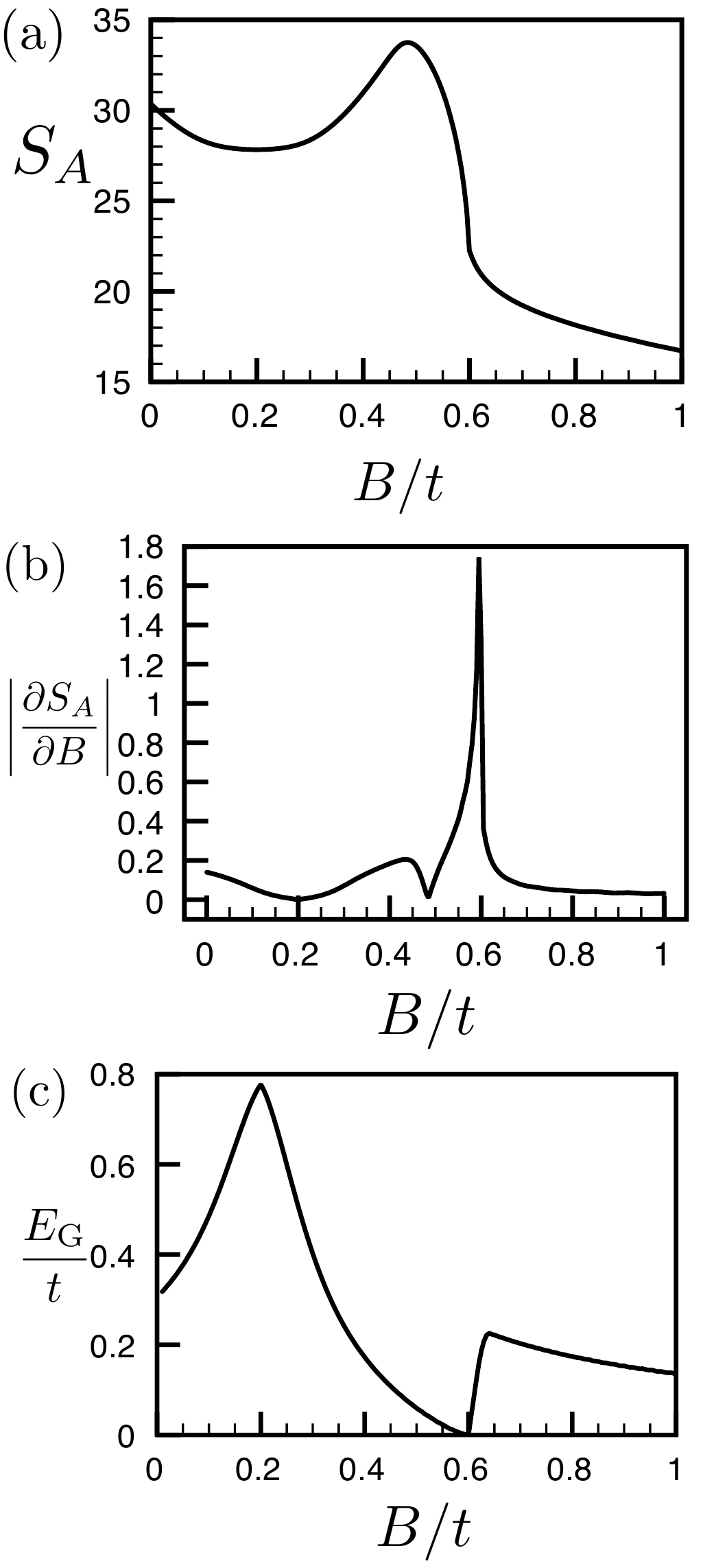}
\caption{{\small
Plot of the entanglement entropy across the phase boundary for the $d$-wave system. Figure (a) shows the entanglement entropy $S_A$ for subregion A a square of side length 12, figure (b) gives $\frac{\partial S_A}{\partial B}$ for the same geometry and figure (c) plots the bulk energy gap as a function of $B$. In the figure we have fixed $\mu=0, A=0.25t, M=0.8t, \Delta_{1}=0.8t$ and $\Delta_{2}=0.4t$.  $B/t=0.6$ is the critical point and $B/t<0.6$ $(B/t>0.6)$ corresponds to the trivial (topological) phase. Notice that the entanglement entropy takes larger values in the trivial phases. This result is different than that of the s-wave topological superconductor shown in Fig.\ref{fig:swavepd}.
     }
     }\label{fig:cut}
\end{figure}

\begin{figure}[t]
\begin{center}$
\begin{array}{cc}
\includegraphics[scale=.22]{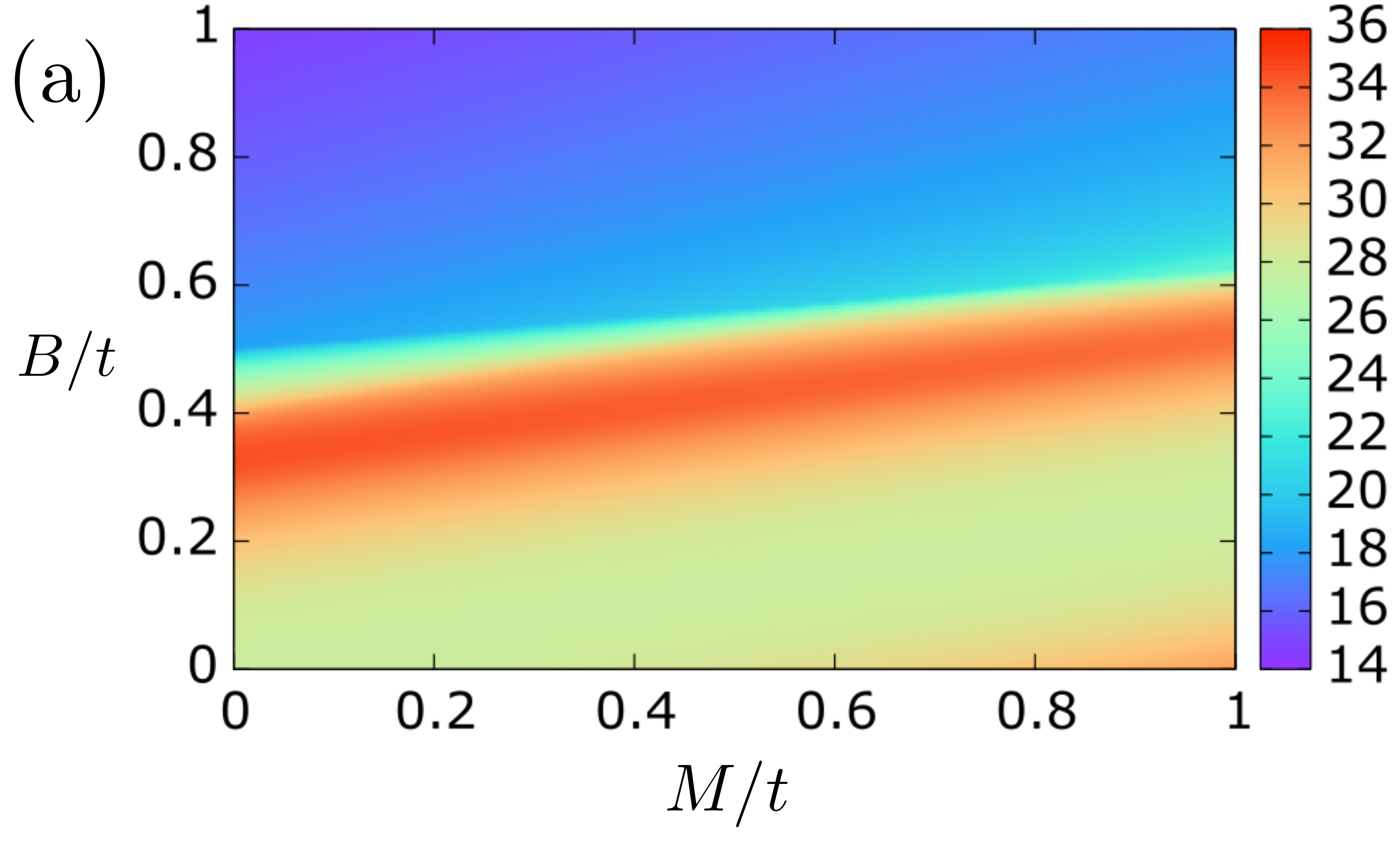} \\
\includegraphics[scale=.22]{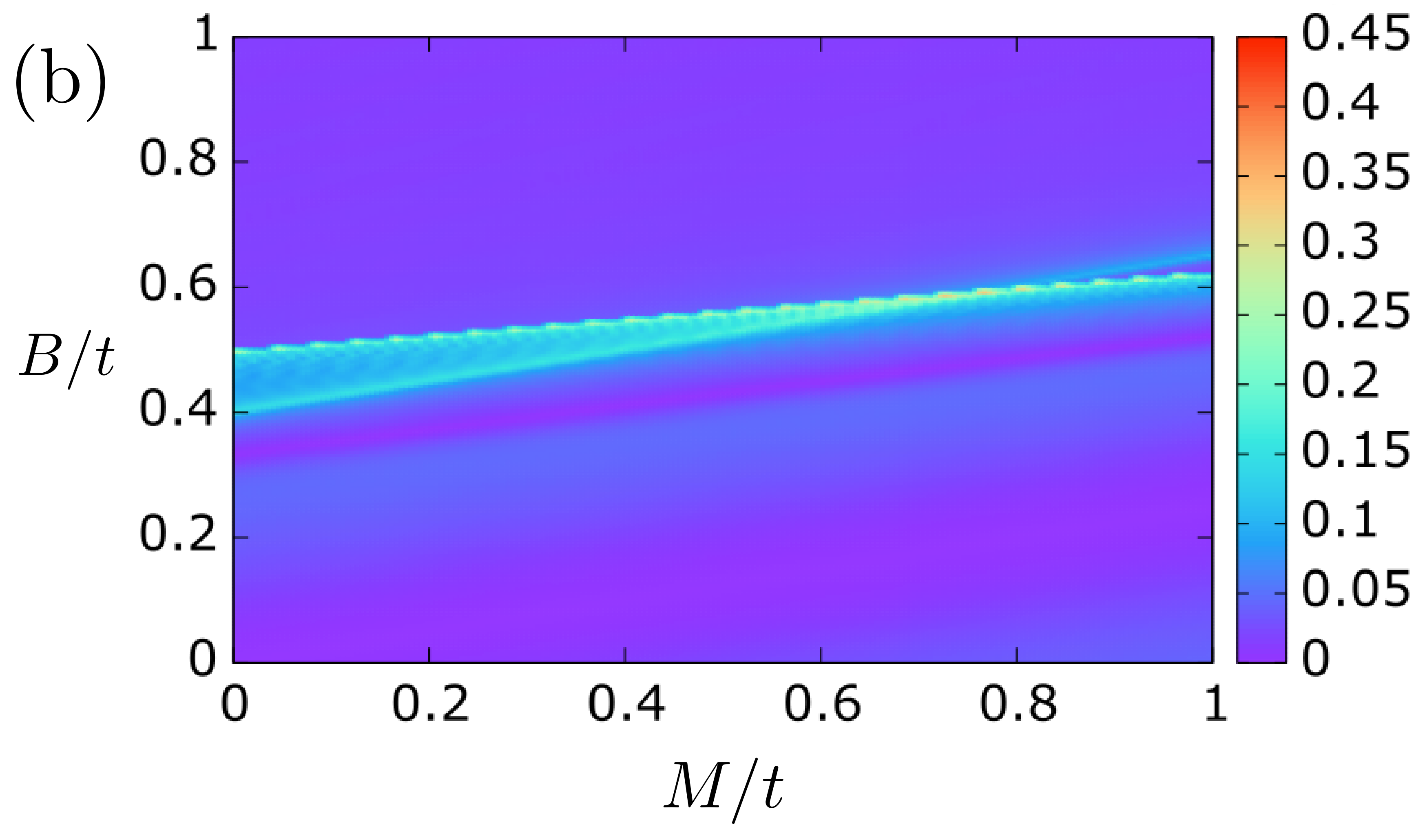} \\
\includegraphics[scale=.22]{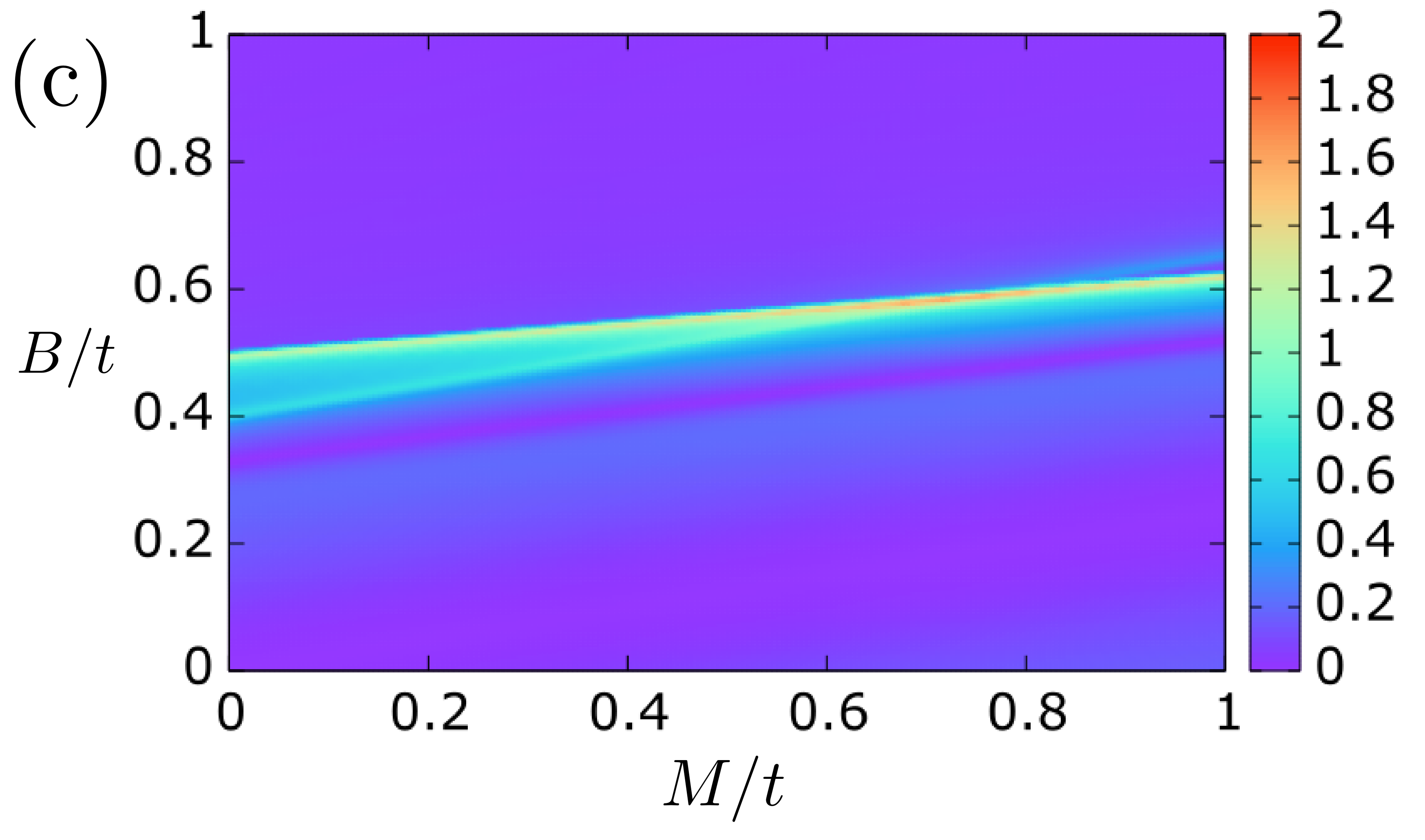} \\
\end{array}$
\end{center}

\caption{{
(Color online) Plot of the entanglement entropy for the $d$-wave system} and its derivatives in $MB$ space. We have (a) the entropy, $S_A$, (b) its derivative $\left|\frac{\partial S_A}{\partial M}\right|$, and (c) the derivative $\left|\frac{\partial S_A}{\partial B}\right|$ . In all figures we have picked a subregion A a square of side length 20 and fixed $\mu=0, A=0.25t, \Delta_{1}=0.8t$ and $\Delta_{2}=0.4t$. The critical line is $B/t=(M/8t+0.5)$ and  $B/t<(M/8t+0.5)$ $(B/t>(M/8t+0.5))$ corresponds to the trivial (topological) phase.
     }\label{fig:dwavepd}
\end{figure}

%One intriguing observation\cite{oliveira} that has been made rather recently in the literature of a spin-orbit coupled {\em triplet} superconductor is that one can see significantly different behaviors of the entanglement entropy depending on which topological phase the model is in. When one proceeds to make a similar observation in our system of interest the conclusion is equally as satisfying. We have found that
When plotting the entanglement entropy and its derivatives with respect to the model's spin-orbit coupling parameters we see the following intriguing property. The topological phase boundaries of our model coincide with ``kinks" in the entanglement entropy. That is, there's a change in behavior of the entanglement entropy at the transition from a trivial superconductor to a topological superconductor.  These kinks are seen as a strong peak in the derivative of the entanglement entropy with respect to material parameters. To make a rather loose analogy with standard thermodynamic variables, the transition appears to be a second order phase transition. A similar property was found in a spin-orbit coupled triplet superconductor in Ref.~[\onlinecite{Oliveira}].

In general, phase transitions between states of different topology but the same symmetry are not characterized by an order parameter.  The entanglement entropy in this case serves as a substitute to a thermodynamic potential and exhibits a kink at the transition.  One may expect that exactly at the transition the bulk gap should close, giving rise to that kink.

\begin{figure*}[t]
  \setlength{\unitlength}{1mm}

   \includegraphics[scale=.45]{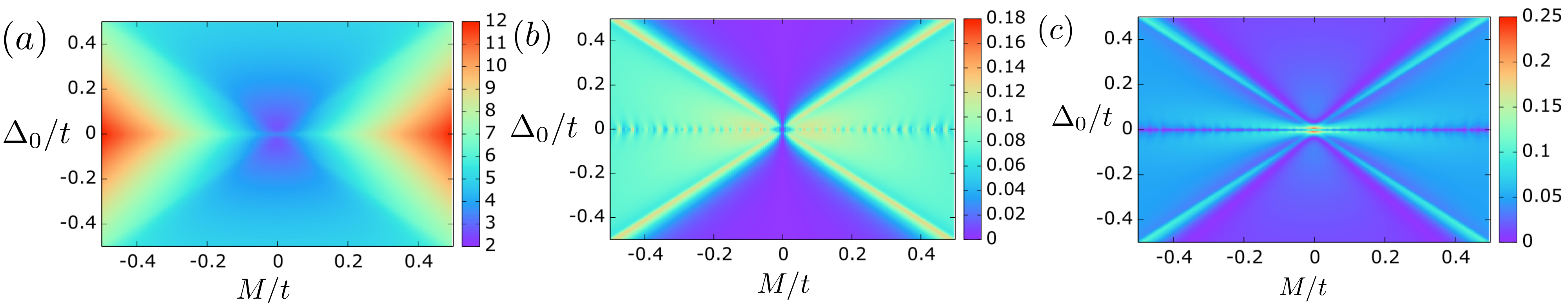}
\caption{{\small
(Color online) Plot of the entanglement entropy and its derivatives in $M-\Delta_0$-space for an $s$-wave system. From left to right we have the entropy, (a) $S_A$, (b) $\left|\frac{\partial S_A}{\partial M}\right|$, and (c) $\left|\frac{\partial S_A}{\partial \Delta_0}\right|$ . In all figures we have picked a subregion A a square of side length 20 and fixed $\mu=-4t, A=0.25t$ and $B=0$. $|\Delta_0|=|M|$ is the phase transition line and $|\Delta_0|>|M|$ ($|\Delta_0|<|M|$) corresponds to the trivial (topological) phase. Notice that the entanglement entropy take larger values in the topological phase, which is opposite to the $d$-wave case.
     }
     }\label{fig:swavepd}
\end{figure*}

%In order to introduce the reader to what we have discussed so far in this section, it is useful to present the entanglement entropy along a cut in parameter space of our model. We present such a cut
In Fig.~\ref{fig:cut} we present a cut through the phase diagram, where only the spin-orbit coupling parameter $B$ is changed. In panel (a) we see that the behavior of the entanglement entropy changes abruptly at $B=0.6t$. This change is more apparent in the derivative of $S_A$ in panel (b).
Checking with the Chern number calculated above, we expect a topological phase transition for this choice of parameters at $B=0.6t$; precisely where this peak occurs. $B<0.6t$ and $B>0.6t$ correspond to the trivial and the topological phases respectively. One might expect that a trivial phase has a smaller value of the entanglement entropy than that of a topological phase because of the absence of the mid-gap entanglement states. However, the entanglement entropy of the d-wave superconductor shows the opposite result; the trivial phase has a larger value of the entanglement entropy. This suggests that in general the leading term of the entanglement entropy cannot be used alone to distinguish trivial phases from topological phases.  However, it does change abruptly at the transition.
In Fig. \ref{fig:cut}c we have plotted the bulk gap of our {\it full} (unpartitioned) system. The most noticeable feature of the gap is that it closes at $B=0.6t$, as is necessary for a topological phase transition.  One may also note that the maximum value of $S_A$ occurs around $B=0.48t$. While we are presently not certain about the origin of this maximum, we may speculate that it is related to some correlation length increase which approaches the system size at $B=0.48t$, before the true transition at $B=0.6t$.

To further explore this behavior we plot $S_A$ and its relevant partial derivatives in parameter space and compare its behavior to the expected phase boundaries. First we explore this for a $d$-wave superconductor. We fix $\mu=0, A=0.25t, \Delta_{1}=0.8t$ and $\Delta_{2}=0.4t$ and explore $M-B$ space. For this specific choice of parameters and focusing on positive values of $B$, we expect a topological phase boundary along the line $B/t=M/8t+0.5$. We have generated data for $S_A$ , $\frac{\partial S_A}{\partial M}$, and $\frac{\partial S_A}{\partial B}$ for this particular choice of parameters, these are presented in Fig.~\ref{fig:dwavepd}.

Studying Fig.~\ref{fig:dwavepd}a, we see a fundamental change in the behavior of the entanglement entropy across the phase boundary line $B/t=M/8t+0.5$.
%$B=t/2 +M/8$
The entropy is large in the trivial phase $(B/t<(M/8t+0.5))$ and then decreases to a lower and much slower changing value across the phase boundary line. This sudden change is more transparent in the derivatives of the entanglement entropy as panels \ref{fig:dwavepd}b and \ref{fig:dwavepd}c.
%%%%%%%%%%%%%%%%%%%%
We see in both of these figures that the derivatives are comparatively small away from the phase boundary lines and increase substantially as these critical points are approached. The exact position of the peak in the derivatives is better seen in the $B$ derivative, as the phase boundary is rather shallow along lines of fixed $B$ which limits our resolution in the $M$ derivative data. Focusing on the plot of  $\frac{\partial S_A}{\partial B}$, one can see a line that is formed by looking for the maximum value of $\frac{\partial S_A}{\partial B}$ for a given value of $M$. Fitting this line gives, to 3 decimal places, a slope of 0.125 and an intercept of 0.500$t$, providing a rather convincing case that $\frac{\partial S_A}{\partial B}$ is peaked along the line $B/t=M/8t+0.5$.

To further study these peaks and also to provide evidence that this behaviour isn't unique to the $d$-wave system, we have also studied the parameter space dependence of $S_A$ in a system with $s$-wave superconductivity. Here we have chosen parameters such that we make as close a connection as possible with the model of Sau {\it et al} in Ref.~[\onlinecite{Sau}].
%In this work the authors are concerned with inducing topological superconductivity in a semi-conductor with Rashba spin-orbit coupling when this quantum well is placed in proximity to a ferromagnetic insulator and an $s$-wave superconductor. Letting $\Delta_0$ be the order parameter of the superconductor, $\mu$ the chemical potential of the quantum well and $V_z$ the Zeeman field produced by the ferromagnet this group finds that provided $\Delta_0^2+\mu^2< V_z^2$  the system will exhibit topological superconductivity.   To make connection with these results in our lattice model
We therefore set $B=0$ and define $\tilde{\mu}=\mu+4t$. In this case our model reduces to that of Ref.~[\onlinecite{Sau}] when the continuum limit is taken.

Using $B=0$, $\tilde{\mu}=\mu+4t$, Eq.~(\ref{chern1}) and assuming $64t^2>- \Delta_0^2+M^2-\tilde{\mu}^2+16t\tilde{\mu} $, the Chern number is simplified to
\beq
C_{1,s} = \frac{\log(\text{sgn}\left[\Delta_0^2 +\tilde{\mu}^2-M^2\right])}{i\pi},
\eeq
where the subscript $s$ denotes $s$-wave. It then follows that if $\Delta_0^2 +\tilde{\mu}^2-M^2<0$ the system is topological. Thus the topological phase boundary is defined by the equation $\Delta_0^2 +\tilde{\mu}^2=M^2$.

We choose to fix $\tilde{\mu}=0$ and study the resulting behavior in the $M$-$\Delta_0$ plane. According to the Chern number we should see phase boundaries at $\Delta_0 = \pm |M|$. Indeed, we see strong indications of a phase boundary along this line. This behavior isn't overtly obvious in the entanglement entropy in Fig.~\ref{fig:swavepd}a, however upon taking derivatives of the data with respect to $M$ and $\Delta_0$ it becomes more apparent. This can be seen in Figs.~\ref{fig:swavepd}b and \ref{fig:swavepd}c, where strong peaks appear along the lines $\Delta_0 =  M$ and $\Delta_0 =  -M$. Thus we have a second clear indiction that $S_A$ changes its behavior across topological phase transition. Comparing to Fig.~\ref{fig:dwavepd}, this demonstration has come from not only a different order parameter symmetry but also from varying a different parameter.

\section{Functional dependence of the entanglement entropy}\label{sec:area}

The study of the functional dependence of the entanglement entropy $S_{A}$ on the `surface area' of a partition $A$ enables one to make conclusions about the ground state of the system. Deviations from the area law have been studied extensively for a variety of different models in different dimensions (see [\onlinecite{eisert}] for a review) and depend on the particular model and ground state under investigation. An example of this in two-dimensional fermionic models can be found in Ref.~[\onlinecite{metlitski}]. This work shows that in models with a spontaneously broken continuous symmetry, the Goldstone mode causes the entanglement entropy to have a sub-leading corner correction proportional to $\ln{L}$, where $L$ is the circumference of the partition. Additionally, for two-dimensional critical fermionic models, one also expects a logarithmic term, not associated with corners\cite{wolf}.

%edited slightly
One difficulty in analyzing the area law is that the circumference of the partition in a lattice model is not uniquely defined. In our calculations we chose the boundary as the line that divides the distance between the outer layer of the partition and the first layer of the complement into half. This is a natural definition as every single lattice point in a line will contribute evenly to the circumference. Other definitions are possible, however, the particular choice should not affect the qualitative behavior of the area law slope, $\alpha$.
\begin{figure}
\begin{center}
$\begin{array}{ccc}
\includegraphics[scale=.33]{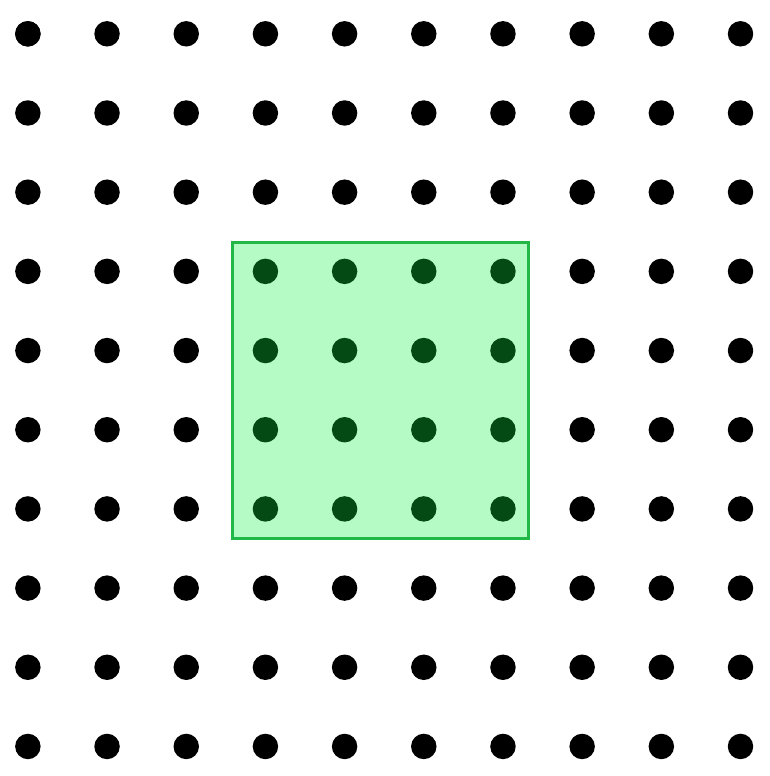} &
\includegraphics[scale=.33]{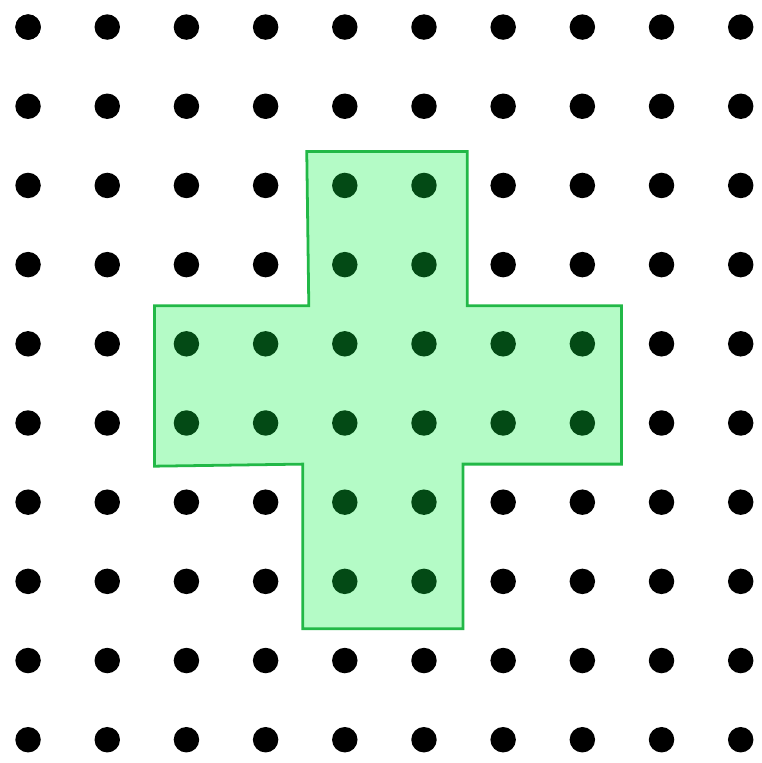} &
\includegraphics[scale=.33]{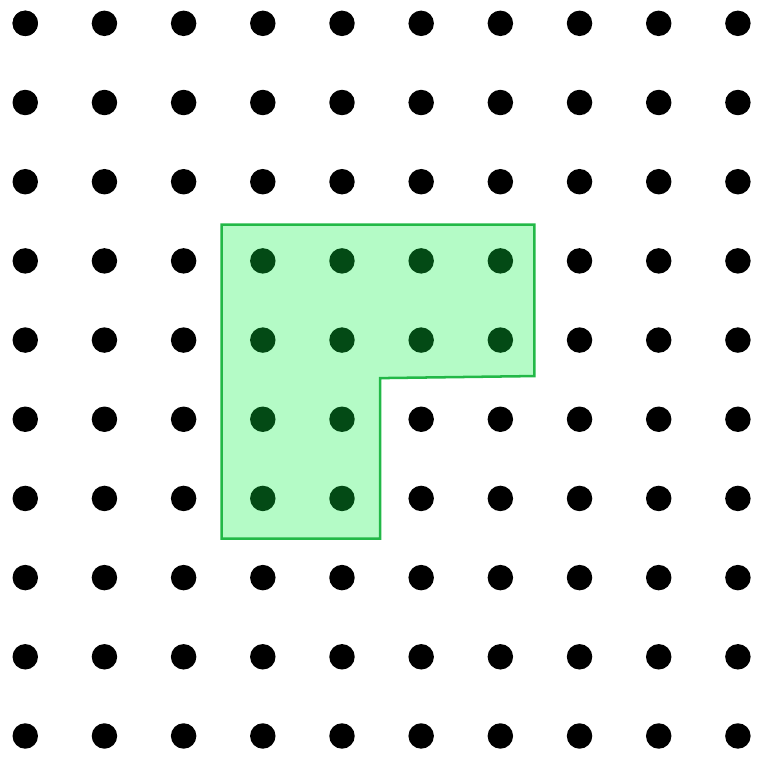} \\
\end{array}$
\includegraphics[scale=.67]{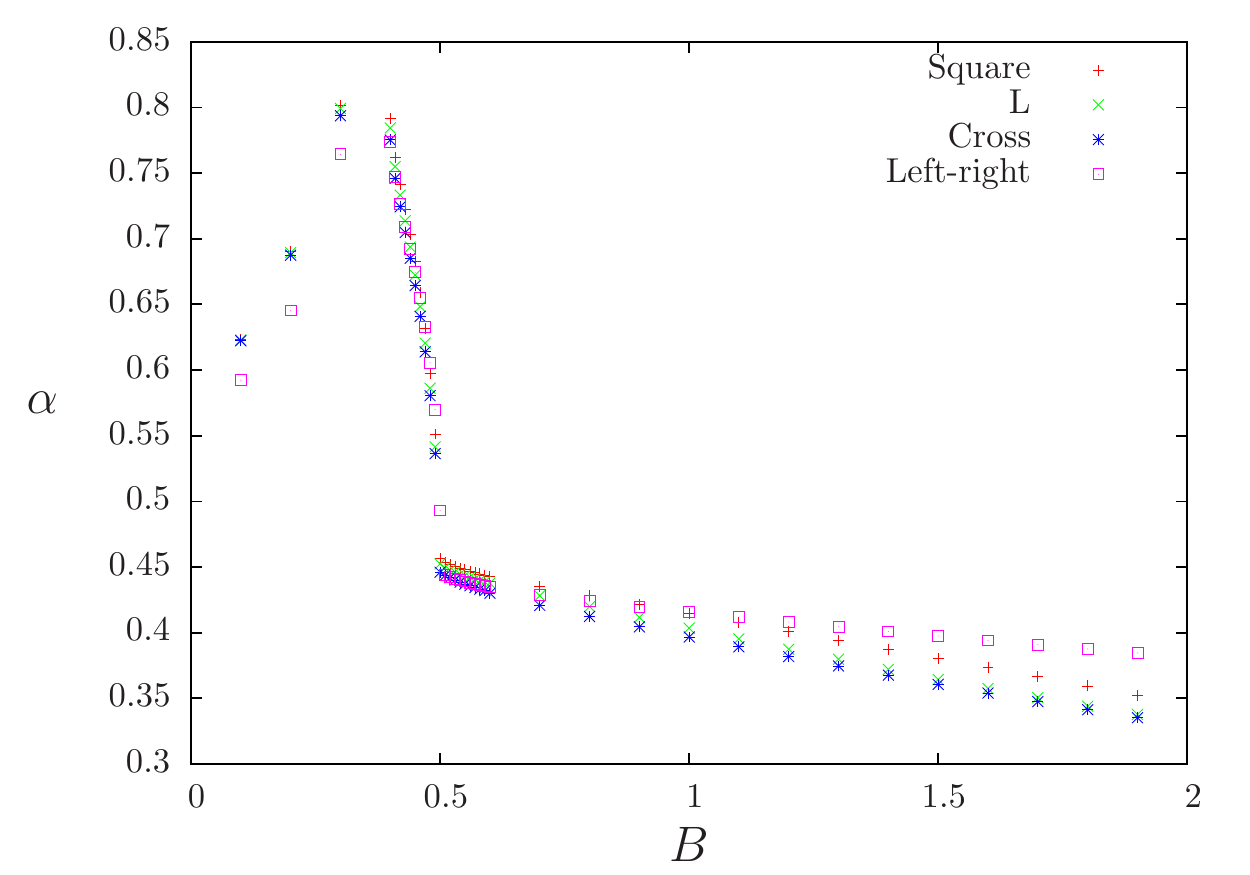}
\end{center}
\caption{(Color online) Upper panel: Schematic plot of the different shapes used for the partition of the system. From left to right: square, cross and (reflected) L-shape. Lower panel: Plot of the linear coefficient in the entanglement entropy for $d$-wave coupling for $M=0$, $\mu=0$, $\Delta_1=0.8t$, $\Delta_2=0.4t$,  and $A=0.25t$ by varying $B$ for a square (red plus), an L-shaped (green cross), a cross shaped partition (blue star) as well as the left right partition (pink square). For this $M$-value the critical point is at $B=0.5$.}\label{fig:alpha}
\end{figure}

\begin{figure}[t]
\begin{center}
\subfloat[~]{\includegraphics[scale=.67]{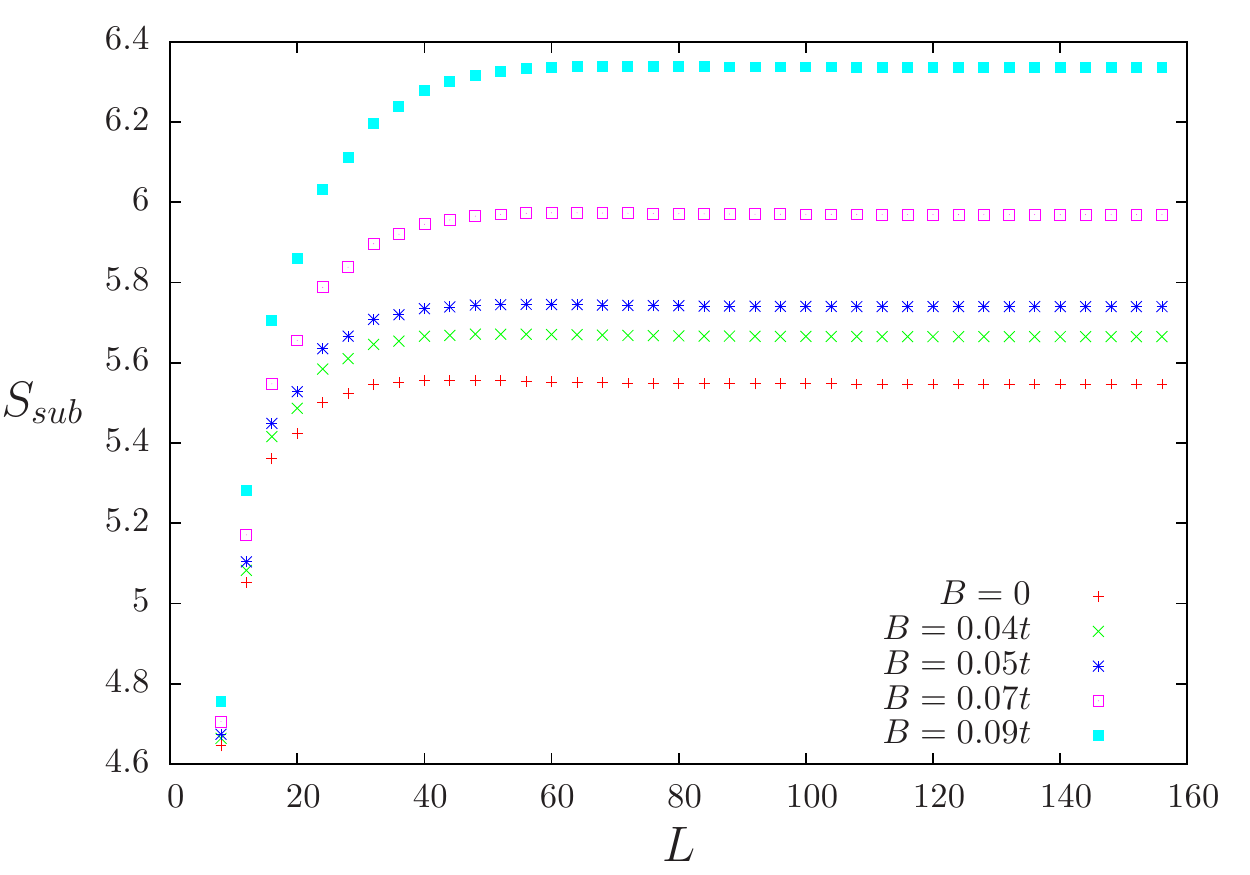}} \\
\subfloat[~]{\includegraphics[scale=.67]{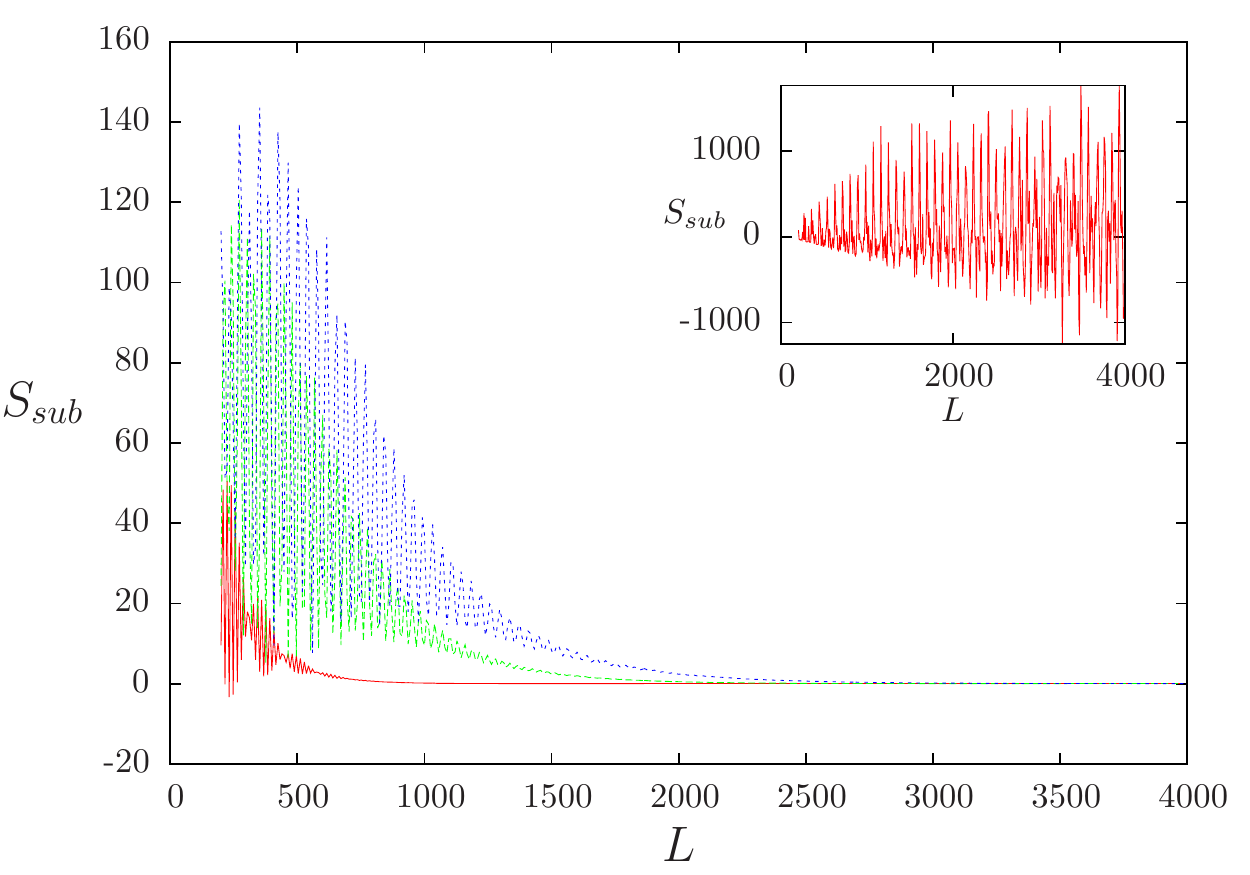}}
\end{center}

\caption{(Color online) Subleading dependence of a $d$-wave superconductor on partition size. (a) $S_{\text{sub}}$ of a square partition as a function of $L$ in the trivial phase for different $B$-values: $B=0$ (red plus), $B=0.04t$ (green cross), $B=0.05t$ (blue star), $B=0.07t$ (pink square) and $B=0.09t$ (cyan diamond). (b) $S_{\text{sub}}$ of a left/right partition in the topological phase in a for varying $B$-values: $B=0.7t$ (red solid line), $B=1.2t$ (green dashed line), $B=1.4t$ (blue dashed line). The inset shows $S_{\text{sub}}$ at the critical point $B=0.5t$. The remaining parameters are fixed at $A=0.25t$, $\mu = 0$, $\Delta_{1}=0.8t$, $\Delta_{2}=0.4t$, and $M=0$.}\label{fig:s_sub}
\end{figure}

For all cases studied in this paper, the leading behavior of the entanglement entropy is linear. The coefficient of the linear term is a non-universal constant denoted by $\alpha$ and dependent on the parameters of the Hamiltonian. In Fig.~\ref{fig:alpha} we plot $\alpha$ for $d$-wave coupling for several $B$-values and different shapes of the partition. Directly at the critical point, the value of $\alpha$ jumps whereas in the topological phase, the change is rather small. Thus, we find that the very distinct signature of the phase transition described in the preceding section is due to the change in $\alpha$. In addition, the dependence of $\alpha$ on the partition shape is rather small and does not change the qualitative behavior. Only for very large $B$-values, a slight difference can be seen.

As the sub-leading nature of these corrections makes it very hard to see them directly in the entanglement entropy, we look at the quantity\cite{ding}.
\begin{align}
\begin{split}
S_{\text{sub}}(L) = LS_{L+1} - (L+1)S_{L},
\end{split}
\end{align}
in which the leading linear term is eliminated.
In the case of only a constant sub-leading term and in the limit of large $L$, $S_{\text{sub}} \propto \text{const}$. For a logarithmic term, the behavior is $S_{\text{sub}} \propto \ln{L}$, whereas for a power law we have $S_{\text{sub}} \propto L^{\eta}$ for some exponent $\eta$. We will also study the dependence of the entanglement entropy on the geometry of the partition. To this end, we will look at a square partition, a cross shaped partition and an L-shaped partition (see the upper panel of Fig.~\ref{fig:alpha}).

In Fig.~\ref{fig:s_sub}(a) we show results for $S_{\text{sub}}$ of a square partition in the trivial phase for the $d$-wave case. It can clearly be seen that in the large $L$ limit, $S_{\text{sub}}$ converges to a constant value, which indicates a constant negative correction to the entanglement entropy. The constant (independent of $L$) it changes with the model parameters. In order to further understand the constant sub-leading term we study differently shaped partitions, such as a cross or an L-shaped partition ({\it c.f.} Fig~\ref{fig:alpha}). As suggested earlier\cite{papanikolaou}, this constant is an effect of the corners, where the dimensions of the partition are of the order of the correlation length. Thus, we would expect to find a constant ratio of the constant of a cross (L-shaped) partition with the constant of a square partition to be 3 (1.5). And indeed, throughout the trivial phase (far away from the critical point), we find the ratios of the constants to be $c_{\text{cross}}/c_{\text{square}} \approx 3 $, and $c_{\text{cross}}/c_{\text{L}} \approx 1.5 $, as expected for a system with zero topological entanglement entropy, $\gamma$. The topological phase, unfortunately, is not reachable in this approach due to finite size effects.

Near a topological phase boundary one must exercise caution when analyzing the functional dependence of $S_A$ on the system boundary size $L$. As the system nears the phase boundary the correlation length grows and so finite size effects become very large. For partitions such as those in Fig.~\ref{fig:alpha}, these finite size effects become important as we are technically limited to modest sized subsystems by the computational time and memory required to diagonalize the matrix $G$ in subsystem A. Using a reasonable amount of memory limits our system size to a side length of $50-60$. Thus, when the correlation length is large we do not have the ability to make our subsystem large enough to see the finite size effects subside. If one is not careful one could misinterpret the finite size effects in this region as some sort of non-trivial subleading contribution to $S_A$, such as $\log{L}$ or $L^{\eta}$.
%deleted red text above

To further illustrate our observation that any subleading terms to $S_A$ for our system originate from corners and at the same time show just how important finite size effects become with an increased correlation length, we use a `corner-less' partition, where subsystem $A$ is a ring on our torus.  If the torus dimensions are $L\times L_l$ where $L_l$ is the longer dimension wrapped around the doughnut hole then our ring dimensions are $L\times l$ and we take $l=L/4$.
%We obtain such a partition by dividing our system into two fractions by a vertical line. We then call the subsystem to the left of this line subsystem A and calculate its entanglement entropy. The top and bottom parts of A are periodically identified on the lattice and so A forms a corner-less subsystem.
The boundary of A is then varied by varying the entire system size.  Besides having no corners this partition has the advantage that translation symmetry along the ring's azimuthal direction is conserved.

Our results for this type of subsystem are illustrated in Fig.~\ref{fig:s_sub}(b). The first striking feature is that $S_{\text{sub}}$ converges to zero for large $L$ for \underline{all} parameter choices. This leads to the conclusion that any subleading terms we have seen above must be a result of corners and subsequently that all subleading behavior beyond the area law for $S_A$ is zero. This is consistent with the observation that the topological entanglement entropy for this system should always be zero.

The second purpose of Fig.~\ref{fig:s_sub}(b) is to illustrate the importance of finite size effects when looking at area laws for spin-singlet superconductors. As the spin-orbit parameter $B$ is increased $S_{\text{sub}}$ acquires a damped oscillatory behavior as a function of $L$. For larger $B$ the amplitude and decay length of these oscillations increase. The way in which $L$ is changed for this partition requires changing both the boundary length of subsystem A {\em and} the total system size. Thus inherent in $S_{\text{sub}}$ are both finite size effects from the fact that $S_A$ depends on the total system size (for smaller lattice sizes, before the thermodynamic limit is reached) and finite size effects from non-area law behavior in $S_A$. The system size required to overcome these effects increases with $B$. We see that even for $B=0.7t$ a very large system size is required before finite size effects vanish. This system size is unreachable using partitions with corners, such as those in  Fig.~\ref{fig:alpha}.

Another indication for finite size effects can be seen in the inset of Fig.~\ref{fig:s_sub}(b), where we show the subleading correction to the entanglement entropy right at the critical point. It displays oscillatory behavior with a very large amplitude which increases with $L$. At this point, the correlation length diverges.

In summary, the current model shows that evaluating the exact subleading dependence of the entanglement entropy on $L$ proves to be far from trivial. This is in contrast to previous work\cite{ding, Oliveira} where such problems did not arise. Therefore, using such subleading terms as a way to evaluate the topology of a specific system (whether they exist or not) may be a prohibitively difficult task. That being said, analyzing the functional dependence on $L$ is not a complete loss in this respect. Looking at the linear coefficient, $\alpha$ the above results suggest that it exhibits a discontinuity at the topological phase boundary.  In Sect. \ref{sec:PBEE} we found that the entanglement entropy is singular at the phase boundary, this could in principle come from a discontinuity in {\it any} term in $S_A$, regardless of the $L$ dependence. The plot in Fig. \ref{fig:alpha} shows that this singularity is in fact coming from $\alpha$. For a finite system with corners we essentially do not know the subleading dependence, whereas in the corner-less partition we find no subleading term at all. Regardless of these two differences we see the same pathological dependence of $\alpha$ on $B$ at the phase boundary.

\section{Edge States in the Entanglement Spectrum}

\begin{figure}
\begin{center}
\includegraphics[scale=.35]{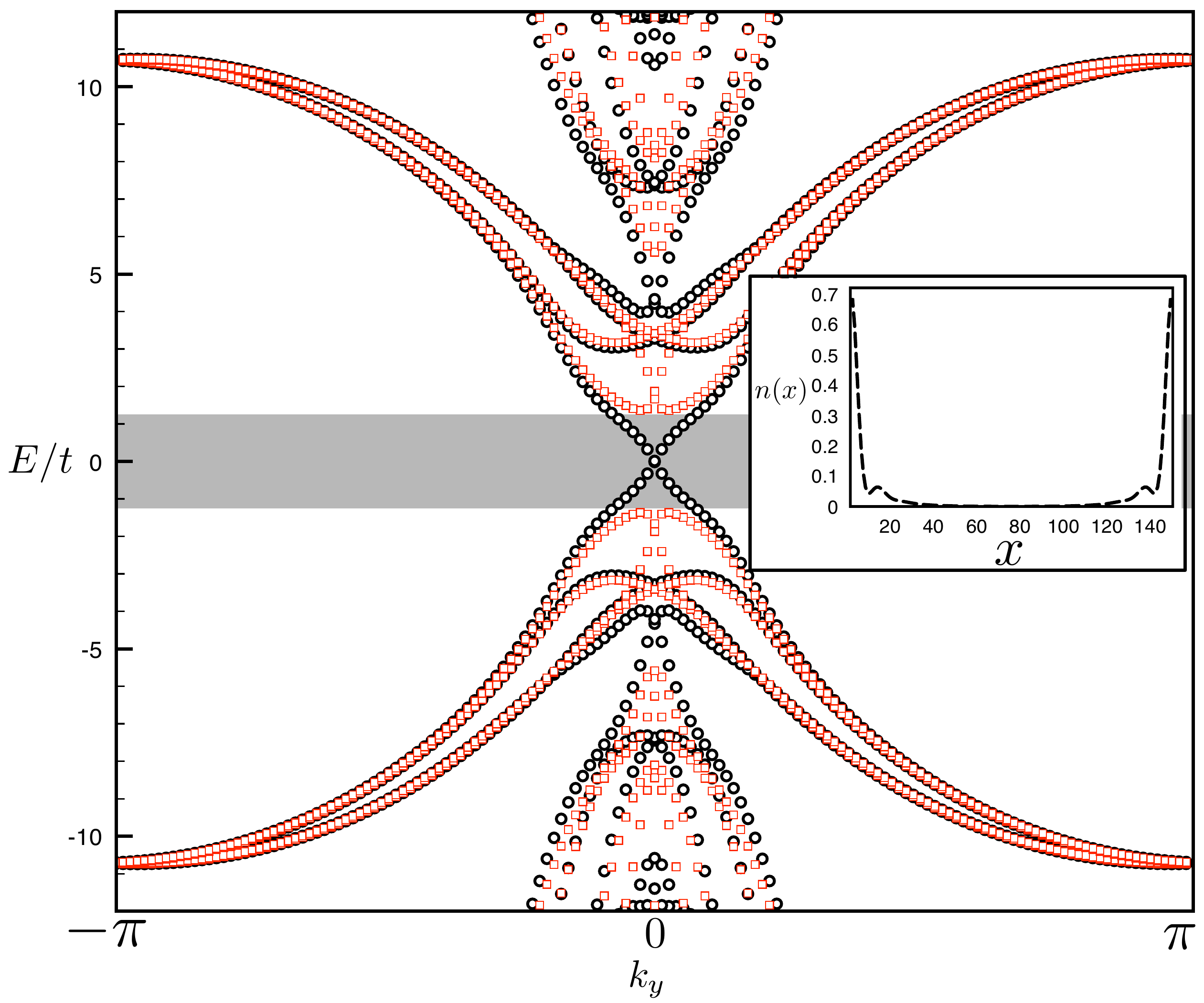}
\end{center}
\caption{(Color online) Entanglement spectrum of an $s$-wave superconductor. Both sets of data are for $\Delta_0=0.3t$, $A=0.25t$, $\mu=-4t$. The red squares are for a system with a $M=0.29t$ (trivial state) while the black circles are from a run with $M=0.31t$ (topological state). The inset shows the density of the zero energy state in the $M=0.31t$ system as a function of position. The boundaries of subsystem A are at $x=0$ and $x=150$ in this inset figure. The gap in the trivial spectrum is shaded to showcase the in-gap states of the topological spectrum.
}
\label{fig:swavespectrum}
\end{figure}

Let us discuss another interesting characteristic of the corner-less partition introduced above. This partition introduces an artificial boundary into the system and therefore we are able to probe boundary physics in a bulk model by looking at the entanglement spectrum of subsystem $A$\cite{li, Qi}. First let us think about a simple $s$-wave model (whose topological phase is a $p$-wave superconductor). If we were to introduce a boundary we would expect to see a zero energy edge mode when the Chern number is 1 and no edge mode when the Chern number is zero\cite{ghosh}.  We can see this same physics in the {\em bulk} model by looking at the entanglement spectrum of the A subsystem. To illustrate this we have plotted the spectrum in the trivial phase and in the topological phase by properly changing parameters. Our results are presented in Fig.~\ref{fig:swavespectrum}. We see quite unmistakably the development of a zero mode upon crossing into the topological region. This zero mode is localized on the boundary of subsystem A, as is shown in the inset of  Fig.~\ref{fig:swavespectrum}.

\begin{figure*}
\begin{center}
\includegraphics[scale=.5]{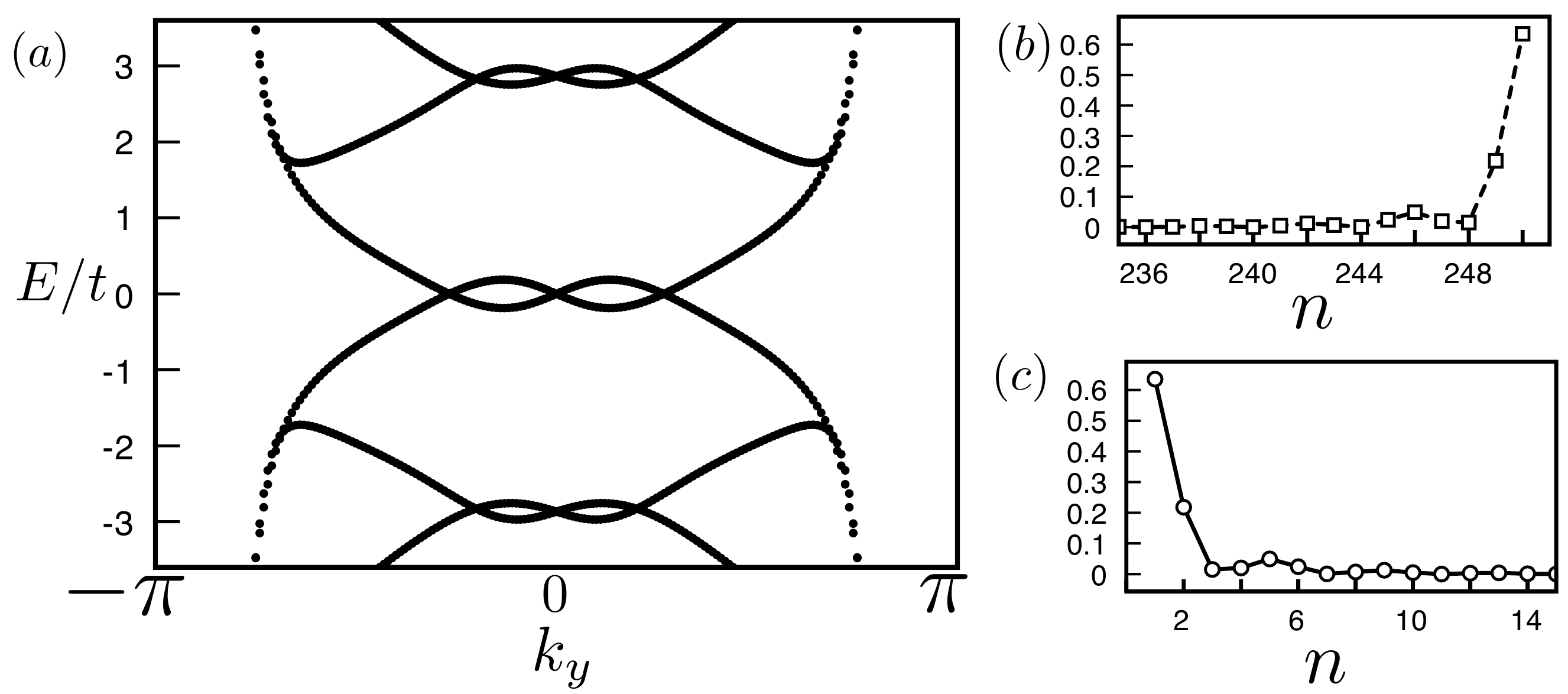}
\end{center}
\caption{Entanglement spectrum and Majorana Modes in a $d$-wave superconductor. This data is for $\Delta_1=0.5t$ ,$\Delta_2=0.8t$, $A=0.5t$, $\mu=-2.5t$ and $M=2t$. (a) The entanglement spectrum;  We note the three zero energy states and therefore the topological nature of the spectra.  (b) and (c)  The probability densities (see Eq. (\ref{densityn})) of the Majorana modes found through orthogonalization as a function of $n$, the number of lattice sites along the direction of the ring which makes up subsystem A.}
\label{fig:dwavespectrum}
\end{figure*}

Finally we explore the edge physics of the {\em bulk} $d$-wave model (whose topological phase is a $p$- or $f$-wave superconductor). The solution of a $d$-wave system with an edge results in a spectrum which is slightly more complicated than the one above for the $s$-wave case\cite{Sato}. From a topological standpoint one expects to see an even number (odd number) of zero energy states when the topology of the system is trivial (non-trivial). We have compared the low-lying states in our corner-less partition entanglement spectrum to those of a physical system with a boundary found, for example, in Reference [\onlinecite{Sato}]. We find consistency between the two with respect to the number of zero energy states, their position in $k$ space as well as their low energy dispersion. A representative example of our results is shown in Fig.~\ref{fig:dwavespectrum}a. Our choice of parameters is such that the low lying states of this plot should be compared with those of Fig.~5II of Ref.~[\onlinecite{Sato}].

Another interesting feature of the data in Figs.~\ref{fig:swavespectrum} and \ref{fig:dwavespectrum}a is the nature of the eigenstate itself, both at and away from zero energy. At lower energies the wave functions are very localized on the edges of the system, with localization length increasing with energy.  This state, however, does not become truly delocalized at any energy.

A second interesting feature of the eigenstates comes from studying the $E=0$ states and, in particular, looking for Majorana modes. We note that it is futile to look for a single Majorana state, as these modes must come in pairs in a finite system. We therefore look for pairs of Majorana states that are spatially separated and reside on opposite sides of the partition.

We notice that the entanglement spectrum exhibits particle hole symmetry, therefore if $|\psi\rangle$ is an eigenstate with energy $E$ then $(\Lambda |\psi\rangle)^*$ is an eigenstate with energy $-E$, where $\Lambda = I \otimes \sigma_x$ with $\sigma_x$ acting on Nambu space and $I$ is the identity on a space of lattice sites and spin.  This leads to the observation that at $E=0$, $|\psi\rangle$ and $(\Lambda |\psi\rangle)^*$ are degenerate eigen states. All eigenstates at $E=0$ are highly localized on the boundary of the system, an example of this is the state plotted in the inset of Fig. \ref{fig:swavespectrum}. Looking for Majorana zero energy states then becomes a task of looking for linear combinations of $|\psi\rangle$ and $(\Lambda |\psi\rangle)^*$ that give states localized at opposite ends of the system and obey the following condition: given two generic linear combinations
\beq
|\phi_{M,i}\rangle = \alpha_{1,i} |\psi\rangle + \alpha_{2,i}  (\Lambda |\psi\rangle)^* = (u_i, v_i)^{T},
\eeq
where $u_i$ and $v_i$ are themselves vectors (each with dimension of one half the dimension of subsystem A) we require $u_i=v_i^*$.

As an example we have studied the gap closure in the $d$-wave spectrum in Fig.~\ref{fig:dwavespectrum}a at $k_y=0$ in detail. Our numerical results give two states with very small energy (approximately $\pm 10^{-12}t$). Treating these two states as degenerate it is possible to form two linear combinations of them which we denote $|M_1\rangle $ and $|M_2\rangle $. We have plotted the density of these states in Figs. \ref{fig:dwavespectrum}b and \ref{fig:dwavespectrum}c. The density of these states on a lattice site $n$ is defined as
\beq\label{densityn}
n_i(n)=|u^{\uparrow}_i(n)|^2+|u^{\downarrow}_i(n)|^2+|v^{\uparrow}_i(n)|^2+|v^{\downarrow}_i(n)|^2,
\eeq
where $u_i(n)$ is the $n^\text{th}$ entry in the vector $u_i$ and is itself a 2-component object (spin-up and spin-down) and the label $i=1,2$ denotes which state we are interested in. Note that this definition is also used in the inset of Fig. \ref{fig:swavespectrum}. The two combinations $|M_1\rangle $ and $|M_2\rangle $ are highly localized on the respective boundaries of subsystem A. Averaging the modulus of the difference between $u_i$ and $v_i^*$ of both of these states over every lattice site (and spin projection) in subsystem A gives a result which is of order $10^{-5}$. Thus these two states are localized on different boundaries and (to a high numerical precision) satisfy the Majorana condition $u_i=v_i^*$.

%changed the section title
\section{Bulk entanglement spectrum and partition induced gap closure}\label{sec:partition}

Looking further into the entanglement spectrum we  note that it is important to specify what kind of partition is used. For example, by partitioning a gapped system into a left and a right part, the low entanglement spectrum is similar to the excitation spectrum near a physical boundary \cite{li,Qi,alba}, as seen above. In fact, states in the ES that are related to bulk degrees of freedom tend to lie very high in the spectrum of such a partition and barely contribute to the entanglement entropy.

Nonetheless, one may extract information about the bulk by defining `extensive partitions', as defined by Hsieh and Fu in Ref.~[\onlinecite{hsiehEE}]. These partitions divide the system into two parts such that the boundary between the two extends throughout the whole system in every direction. Thus, the partition forms a superlattice. The periodicity of an extensive partition removes the edge modes from the ES and lead it to resemble a bulk spectrum.

\begin{figure}[t]
\begin{center}

\subfloat[~]{\includegraphics[scale=.44]{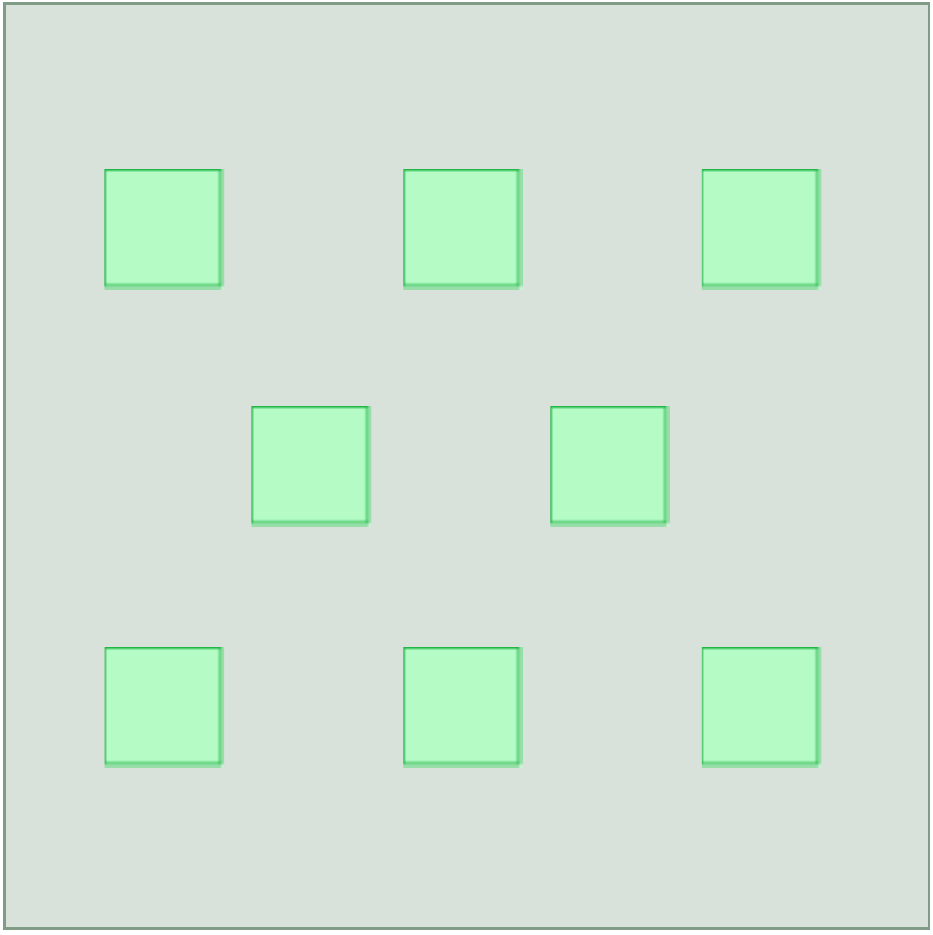}} \quad
\subfloat[~]{\includegraphics[scale=.44]{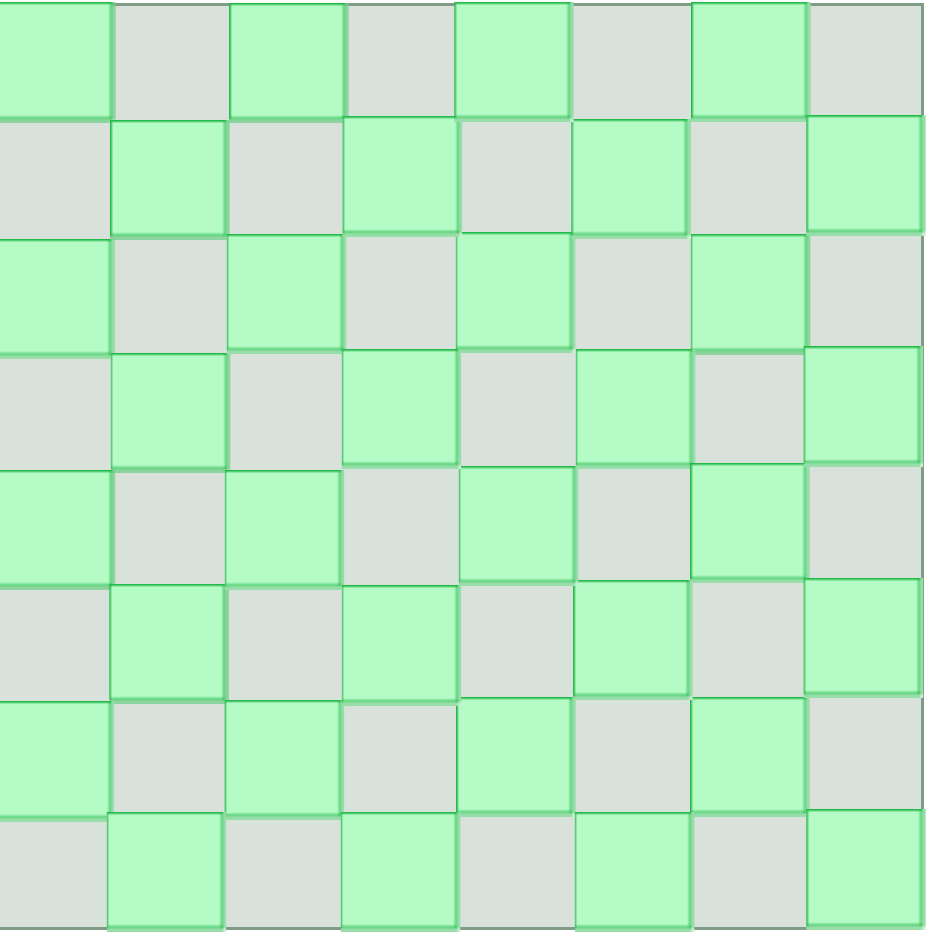}}
\end{center}

\caption{{
(Color online) (a) Schematic display of an asymmetric partition where the green squares are subsystem $A$ and $B$ its compliment. (b) Symmetric partition
     }
     }\label{fig:partlim}
\end{figure}

An example for an extensive partition is show in Fig.(\ref{fig:partlim}) where one subsystem is a collection of square islands while the other is the remaining sea.  Using these partitions we demonstrate that the ES may exhibit a topological phase transition as the partition is changed.  Throughout the following discussion we fix the parameters of the model such that it represents a topological state.  The only thing we change is the partition.  As the result of this change a phase transition appears in the entanglement spectrum while the physical spectrum is always gapped and topological. 

The tuning of partitions is done as follows.  In the beginning system $B$ consists of islands while $A$ is the sea.  In one extreme case the island size is shrunk to zero so that $B$ is an empty set while $A$ is the whole physical system. We then gradually enlarge the islands.  At some point the islands corners touch.  This is called the symmetric point.  When the islands grow further they overlap such that system $B$ becomes the sea and system $A$ breaks into isolated islands.  In Ref.~[\onlinecite{hsiehEE}] Hsieh and Fu argue that in both extremes ($A$ or $B$ being the full system) subsystem $A$ is gapped.  However, when $A$ includes the full system it is in a topological state (like the physical system) but when it is a collection of vanishingly small islands it is connected to the atomic limit (a trivial state).  They conclude and demonstrate on a topological insulator that somewhere between these two limits the ES of subsystem $A$ undergoes a topological phase transition which is manifest an a gap closure.

\begin{figure*}[t]
\begin{center}
\subfloat[~]{\includegraphics[scale=.5]{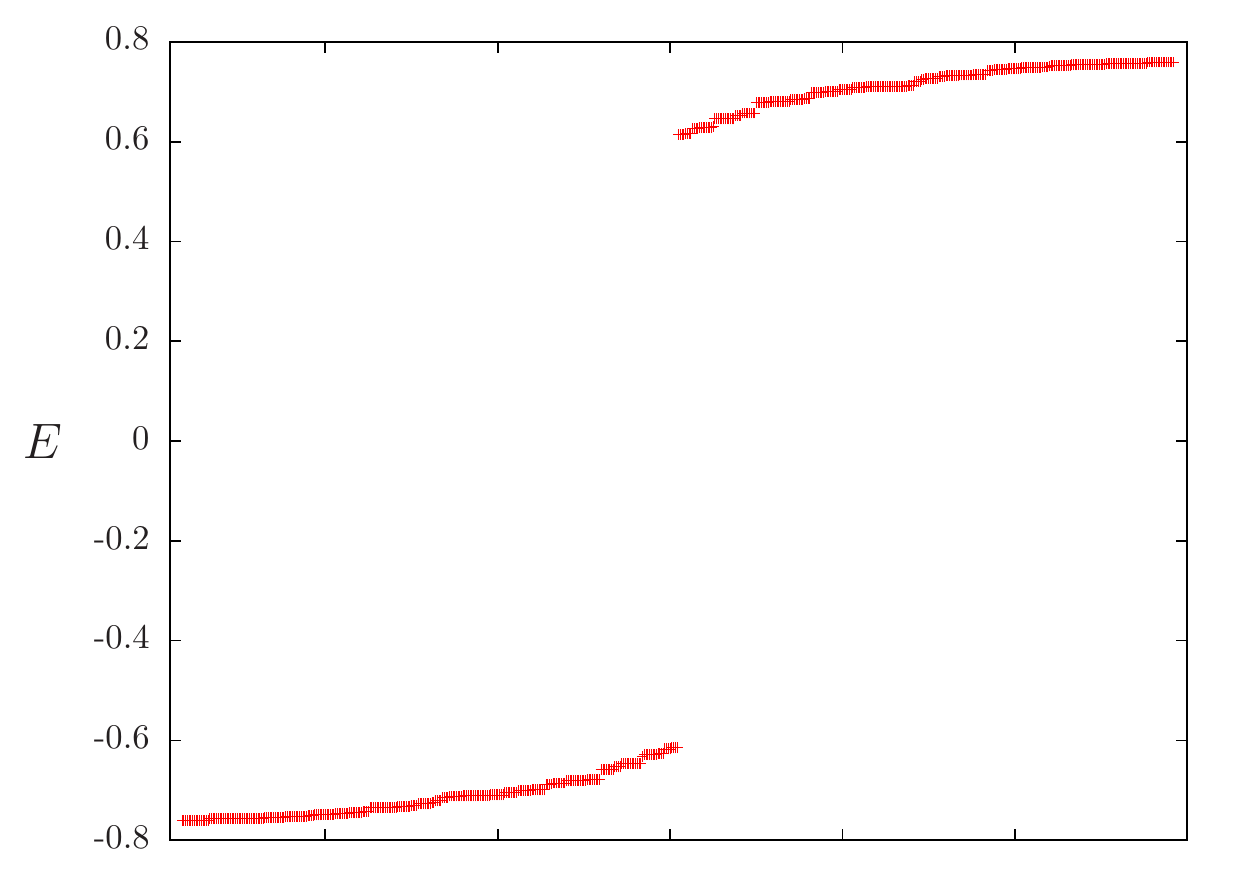}}
\subfloat[~]{\includegraphics[scale=.5]{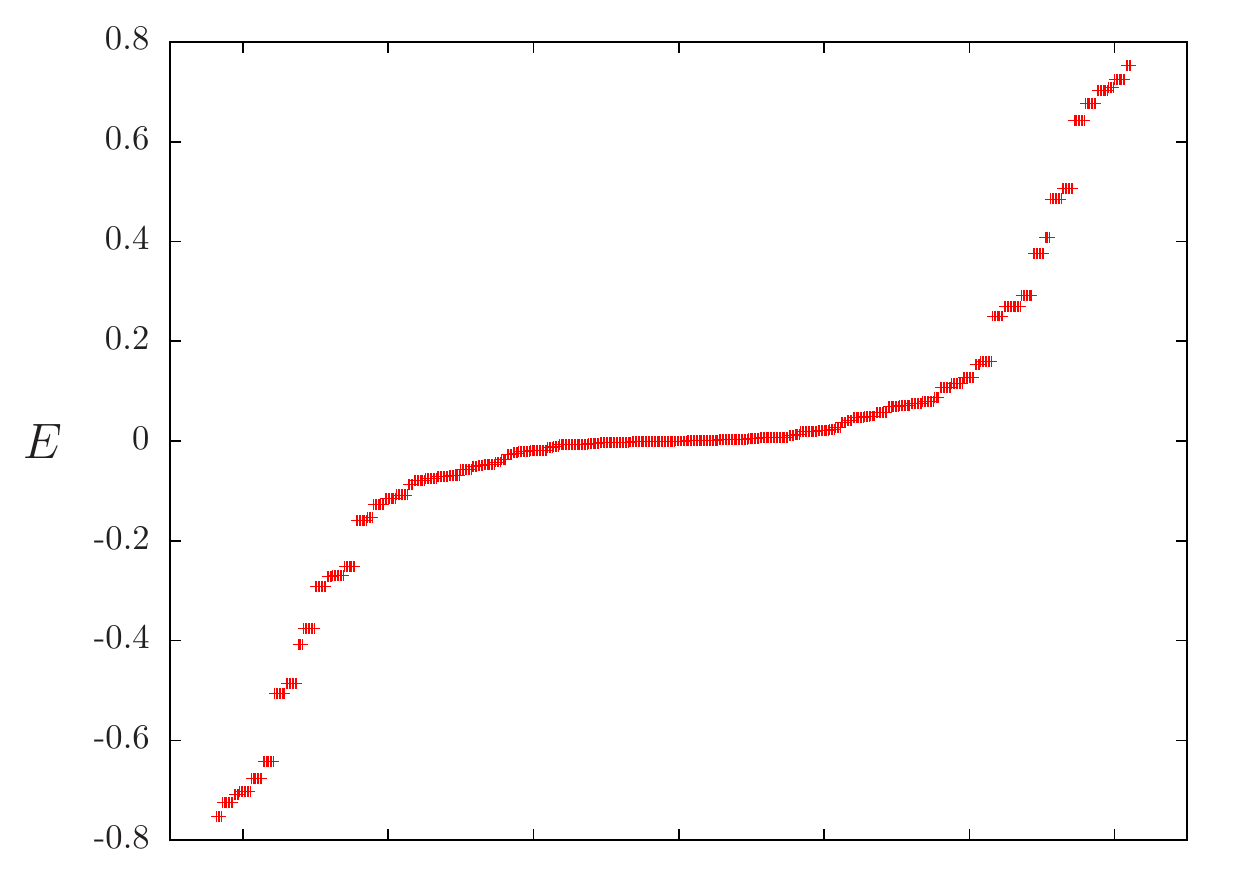}} \quad
\subfloat[~]{\includegraphics[scale=.5]{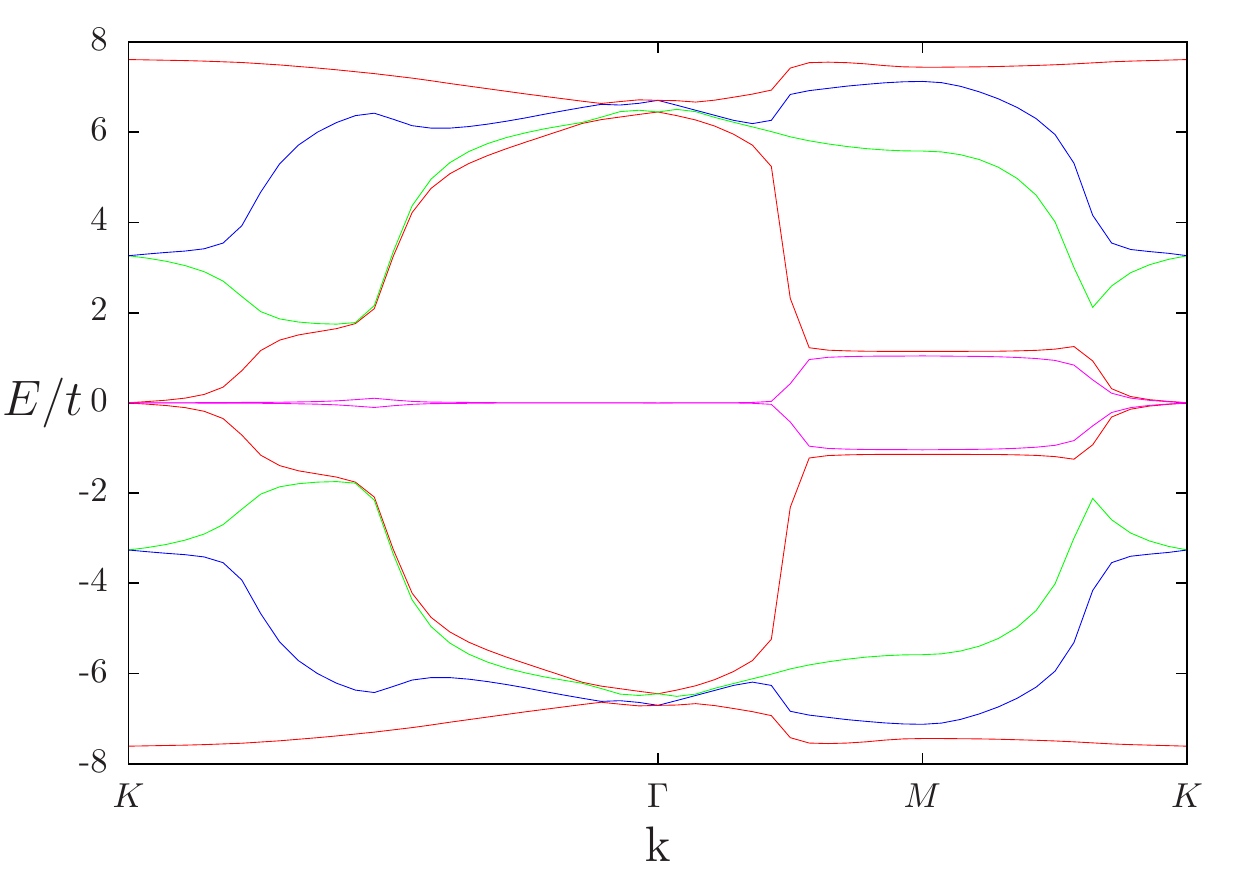}}
\subfloat[~]{\includegraphics[scale=.5]{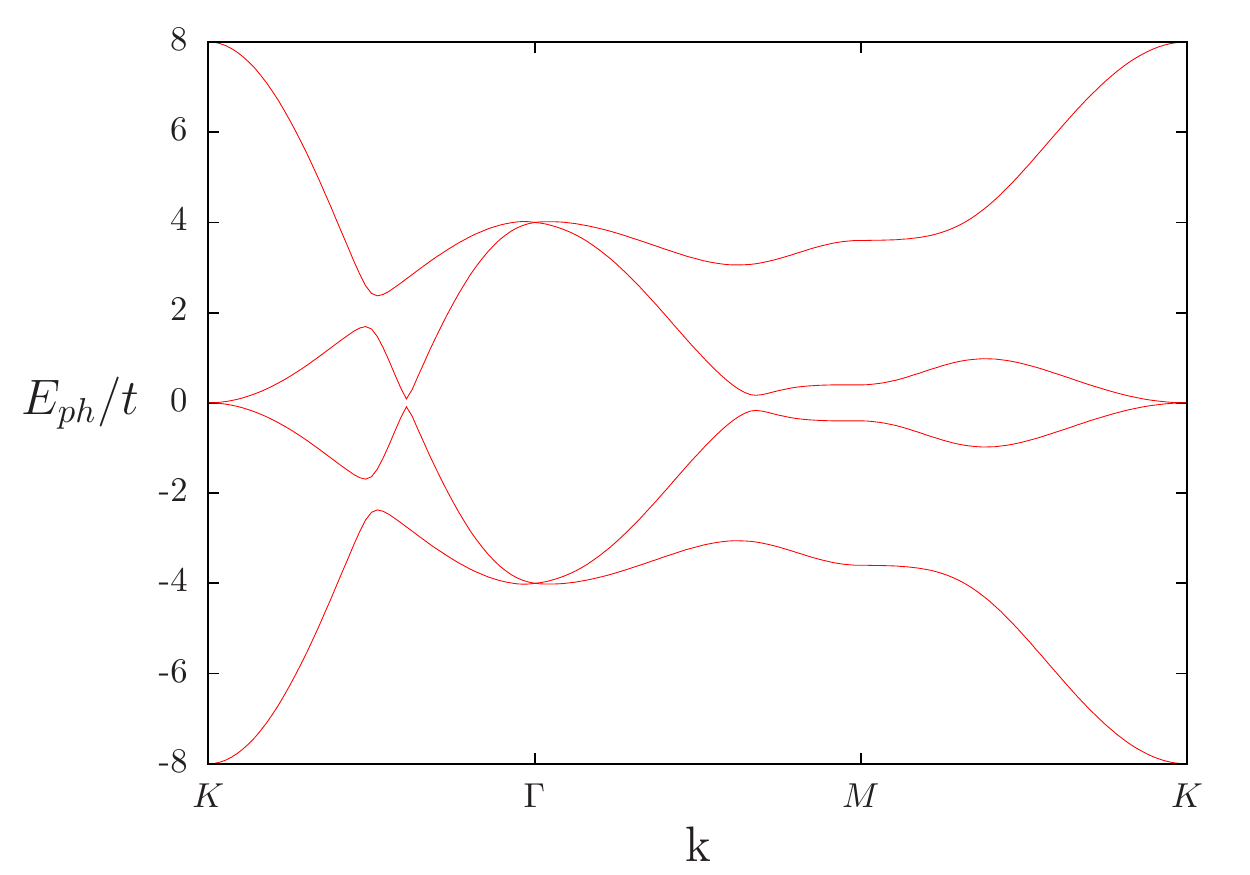}}
\end{center}

\caption{{
(Color online) (a) Low energy part of the entanglement spectrum of the asymmetric partition in the topological phase of a $d$-wave superconductor with $B=0.7t$, $M=0.3t$, $\mu=0$, $\Delta_1=0.8t$, $\Delta_2=0.4t$,  and $A=0.25t$. (b) Low energy part of the entanglement spectrum of the symmetric partition in the topological phase of a $d$-wave superconductor (with the same parameters as in (a)). (c) $k$-space spectrum for the symmetric partition of a 112 by 112 square lattice with the same parameters as in (a), where the $k$-vectors are defined with respect to the superlattice. (d) The physical spectrum of the system when its parameters are tuned to the critical point at $B=0.5t$ and $M=0$.
     }
     }\label{fig:entspec}
\end{figure*}

Following Hsieh and Fu we apply the above idea to a topological superconductor.  In Fig.~\ref{fig:partlim}a, we have sketched an extensive partition while the symmetric point is shown in Fig.~\ref{fig:partlim}b. In both cases, for a $d$-wave as well as an $s$-wave SC, the ES in the asymmetric cases are gapped, as can be seen for the case of a $d$-wave SC in Fig.~\ref{fig:entspec} independent of the phase the system is in. Staying in a topologically non-trivial physical state, we can now induce a phase transition in the entanglement spectrum by varying the partition across the symmetric point. As can be seen in Fig.~\ref{fig:entspec}, this indeed induces a gap closure.

%deleted a red sentence
Using the fact that the symmetric partition forms a superlattice, one can define $k$-vectors with respect to the superlattice and arrange the states in the ES with momentum.  The ES can then be compared to the spectrum of an unpartitioned system, whose parameters are set to the critical point. The result can be seen in the lower part of Fig.~\ref{fig:entspec}.  In Fig.~\ref{fig:entspec}(c) we can see the ES of the symmetric extensive partition  qualitatively mirrors the physical spectrum of a critical, unpartitioned system (Fig.~\ref{fig:entspec}(d)). The parameters for the system in Fig.~\ref{fig:entspec}(c) where chosen to be in the topologically non-trivial phase. Thus, the symmetric partition realizes the critical system \emph{without} changing the model parameters. The gap closes at the $K$-point, where the spectrum has a massless Dirac cone.

\section{Conclusion}\label{sec:conclusion}

In this paper we have studied several proposed signatures of topology in the entanglement entropy and spectrum of superconducting models with topological phases. Our systems of interest are spin-orbit coupled superconductors, motivated by recent progress in the search for Majorana fermions\cite{Farrell1, Sau}. We have compared our results with those obtained in previous work as well as evaluated the potential use of each of the methods for the study of more complicated (disordered/interacting) systems, where the topology is not known {\it a priori}\cite{Progress}.

We have analyzed the dependence of a bipartite partition on the circumference of the partition and found a dependence of the form $S(L) = \alpha L + \dots$, where the first term is the celebrated area law and the dots stand for sub-leading terms. The coefficient $\alpha$ was found to have a sharp kink right at the phase transition such that it captures the transition very clearly.  In the trivial phase, the only sub-leading term was found to be a constant caused by corner effects. Meanwhile, the topological phase is not easily classified using a small finite system (due to finite-size effects)  and we must defer to a corner-less system.  In the corner-less partition the EE is given by the area law without any subleading terms. We conclude that any non-area law contributions in the finite system must be due to corners. As expected, throughout all phases the topological entanglement entropy, $\gamma$, was found to be zero.  Therefore, calculating $\alpha$ for a corner-less partition and looking for singular behaviour may be of potential interest in more complicated systems.

%In this paper we showed the existence of multiple signatures of the topological phase transition in a spin-orbit-coupled d-wave and s-wave superconductor. First of all, we analyzed the dependence of a bipartite partition of the circumference of the partition and found a dependence of the form $S(L) = \alpha L + \dots$, where the dots stand for sub-leading terms. The linear coefficient $\alpha$ was found to have a sharp kink right at the phase transition point such that by varying the physical parameters of our model one could clearly see a transition in the behavior of the entanglement entropy. In the trivial phase, the only sub-leading term was found to be a constant which was determined to be fully caused by corner effects. Throughout all phases, the topological entanglement entropy $\gamma$ was found to be zero.

Another signature of the topology of the system can be found by looking at the entanglement spectrum. Depending on the choice of partitioning one may obtain different topological properties of the entanglement Hamiltonian.  A phase transition between the topological and the trivial phase can be seen as a gap closure in the entanglement spectrum.  This is obtained by changing the extensive partitioning while leaving the physical parameters unchanged.  This property is related to the non-trivial topology of the underlying state.  Moreover, this finding implies that one has to apply special care when using the entanglement spectrum to extract information about the ground state of a physical system as it can undergo a phase transition while the physical system does not.

%deleted red

In addition to our goals stated in the introduction we would also like to emphasize the versatility of the approach outlined in this whole paper; it can be applied to all quadratic models with or without translational invariance where in the latter case the system sizes are limited by computational power. The use of various forms of partitions leads to a consistent picture of the different topological phases of a system, as shown for a spin-orbit coupled superconductor with $d+id$- and $s$-wave coupling.

\section{Acknowledgements}
We are grateful for useful discussions with R.~Melko. Financial support for this work was provided by the NSERC and FQRNT (TPB, JB, SM) the Vanier Canada Graduate Scholarship and the Walter C. Sumner Memorial Fellowship (AF). The majority of the numerical calculations were performed using CLUMEQ/McGill HPC supercomputing resources.

\bibliographystyle{apsrev}
\bibliography{topoSC}

\end{document}